\documentclass{IEEEtran}
\usepackage{graphicx}
\usepackage{booktabs}
\usepackage{multirow}
\usepackage{array}
\usepackage{adjustbox}
\usepackage{float}
\usepackage{placeins}

\usepackage[table]{xcolor}
\definecolor{lightgreen}{RGB}{200,255,200}
\definecolor{tmlcncolor}{cmyk}{0.93,0.59,0.15,0.02}
\definecolor{NavyBlue}{RGB}{0,86,125}
\usepackage{amsmath,amssymb,amsfonts}
\usepackage{amsthm}
\usepackage{mathrsfs}

\usepackage{algorithm}
\usepackage{algpseudocode}

\usepackage{textcomp}
\usepackage{xspace}
\usepackage{url}
\usepackage{comment}
\usepackage{balance}
\usepackage{manyfoot}

\usepackage{cite}
\usepackage{hyperref}
\hypersetup{hidelinks=true}
\usepackage{cleveref}

\usepackage{subcaption}  

\usepackage{tikz}
\usepackage{tikz-dependency}

\usepackage{microtype}

\newcommand*\circled[1]{%
\tikz[baseline=(char.base)]{
\node[shape=circle,fill,inner sep=0.8pt] (char)
{\textcolor{white}{#1}};}}

\theoremstyle{definition}
\newtheorem{definition}{Definition}[section]

\newcommand{\framework}{\textsc{Prometheus}\xspace}
\newcommand{\newmodel}{Lightweight\xspace}

\begin{document}


\title{An Explainable Failure Prediction Framework for Neural Networks in Radio Access Networks }
\author{
\IEEEauthorblockN{Khaleda Papry\IEEEauthorrefmark{1}, Francesco Spinnato\IEEEauthorrefmark{2,3}, Marco Fiore\IEEEauthorrefmark{4}, Mirco Nanni\IEEEauthorrefmark{3}, Israat Haque\IEEEauthorrefmark{1}}

\IEEEauthorblockA{\IEEEauthorrefmark{1}Department of Computer Science, Dalhousie University, Canada}
\IEEEauthorblockA{\IEEEauthorrefmark{2}Department of Computer Science, University of Pisa, Italy}
\IEEEauthorblockA{\IEEEauthorrefmark{3}ISTI CNR, Pisa, Italy}
\IEEEauthorblockA{\IEEEauthorrefmark{4}IMDEA Networks Institute, Madrid, Spain}}

\maketitle
\begin{abstract} As 5G networks continue to evolve to deliver high-speed, low-latency, and reliable communications, ensuring uninterrupted service has become increasingly critical. While millimeter-wave (mmWave) frequencies enable gigabit data rates, they are highly susceptible to environmental factors, often leading to radio link failures (RLF). Predictive models leveraging radio and weather data have been proposed to address this issue; however, many operate as black boxes, offering limited transparency for operational deployment.
This work bridges that gap by introducing \framework, a framework that combines explainability-based feature pruning with model refinement. \framework can be integrated into state-of-the-art predictors such as GNN-Transformer and LSTM-based architectures for RLF prediction, enabling the development of accurate and explainability-guided models in 5G networks. It provides insights into the contribution of input features and the decision-making logic of neural networks, leading to lighter and more scalable models.
When applied to RLF prediction, \framework unveils that weather data contributes minimally to the forecast in extensive real-world datasets, which informs the design of a leaner model with 50\% fewer parameters and improved F1-scores with respect to the state-of-the-art solution. Ultimately, \framework empowers network providers to evaluate and refine their neural network–based prediction models for better interpretability, scalability, and performance.

\end{abstract}



\section{Introduction} \label{introduction}

With the rapid evolution of telecommunications, 5G networks have become a key technology, offering ultra-low latency, high-speed connectivity, and massive device support. By 2030, global 5G subscriptions are projected to reach 6.3 billion, covering about 67\% of mobile users and enabling applications such as autonomous vehicles and emergency response. The required reliability and performance can be achieved using millimeter wave (mmWave) spectrum to deliver gigabit data rates \cite{niu2015survey}. Major carriers (e.g., AT\&T, Orange) are deploying mmWave to enhance Quality of Service (QoS) in urban and rural areas. However, mmWave links are vulnerable to attenuation and distortion, causing radio link failures (RLFs) that disrupt service and increase operational costs \cite{aktacs2022towards}. RLF occurs when signal quality or stability falls below recovery thresholds, resulting in severe degradation or loss of connectivity \cite{ieee-rlf-survey}. Timely and accurate RLF prediction is therefore crucial to maintain network reliability. Data-driven prediction in the 5G Radio Access Network (RAN) can leverage variations in time of key performance indicators (KPIs) as well as the environmental context (e.g., weather).
Timely and accurate RLF prediction is therefore crucial to maintain network reliability. Data-driven prediction in the 5G Radio Access Network (RAN) can leverage both key performance indicators (KPIs) and environmental context. For instance, Islam et al.~\cite{islam2022deep} proposed an Long-Short Term Memory (LSTM) autoencoder, while Hasan et al.~\cite{hasan2023transformer} introduced a Graph Neural Network (GNN) and Transformer-based architecture. These models forecast radio link failures by analyzing historical KPI and weather station data. However, despite good performance, they do not explain the role of weather data or their internal decision processes. This black-box nature limits interpretability, which is essential for network operators to take timely and informed preventive actions~\cite{theissler2022explainable}.

For these reasons, interest in Explainable Artificial Intelligence (XAI) techniques for temporal data is rapidly increasing~\cite{theissler2022explainable}. Such methods enhance transparency and provide valuable insights into machine learning (ML) decision-making by identifying the importance of input features and revealing the rationale behind model predictions~\cite{longo2020explainable}. This is particularly relevant in the time series domain~\cite{refoyo2024sub, spinnato2024fast}, where explainability plays a key role in improving model trustworthiness and supporting practical deployment~\cite{bianchi2024multivariate}.
Classical XAI were recently integrated in mobile networks in recent years, for example, SHapely Additive exPlanations (SHAP) \cite{lundberg2017unified} was deployed in \cite{fiandrino2024aichronolens, gholian2025deexp} to understand root causes of model errors \cite{fiandrino2024aichronolens} or spot vulnerabilities \cite{gholian2025deexp}. Fern\'andez P{\'e}rez et al. \cite{perez2025chronoprof} generalized these forecasting schemes, considering both univariate and multivariate mobile traffic.  
Despite their effectiveness, two main gaps remain~\cite{senevirathna2024survey}: post-hoc XAI analyses have primarily been used for interpretability and debugging but not to inform or guide architectural refinements, leaving a disconnect between model explanations and subsequent model improvements; and existing solutions have not provided systematic assessments or explanatory insights into the effectiveness of state-of-the-art Transformer-based architectures in time series tasks, limiting understanding of how and why these models achieve their performance.

This work fills these gaps by proposing an explainable framework for 5G Radio Access Network (RAN) called \framework. It consists of two fundamental steps: XAI-based model analysis and informed model refinement. In particular, \textit{\textbf{(i)}} \framework adapts SHAP's model-agnostic explanations, to RLF predictions to assess feature importance and understand the decision-making logic, using Monte Carlo approximation. The output contribution scores are then captured in a saliency map and aggregated across samples to generate the global importance of the features across all instances to identify the most important features for the prediction of the model. Then \textit{\textbf{(ii)}}, we revisit the model architecture to remove components that do not contribute to the RLF prediction. 
\framework can be integrated into any complex black-box model. In particular, we implement and extensively evaluate \framework using a GNN-Transformer and an LSTM-based RLF prediction model on a complex real-world telecom operator dataset \cite{aktacs2022towards} both in rural and urban deployments scenarios for a total of 0.4 and 1.8 million data points, respectively, containing radio link time series KPIs as well as surrounding weather station observations. 
In summary, our contributions are the following:

\begin{itemize}
\item We propose \framework, an XAI-driven RLF prediction framework for 5G RAN, enabling model interpretation and integration with state-of-the-art LSTM and Transformer architectures for multivariate time series.
\item We analyze the impact of historical context length and weather data on prediction performance, using XAI-based feature importance to identify and retain only the most relevant features.
\item We apply post-hoc architectural refinement by removing non-contributing components, resulting in a lightweight Transformer with 50\% fewer parameters and improved F1-score, and an LSTM-based model with 92\% fewer parameters and comparable accuracy.
\item We validate \framework on two real-world datasets, demonstrating its effectiveness, scalability, and ease of integration into existing neural network models for practical deployment.
\item We release the \framework artifact to support reproducibility and future research.
\end{itemize}

The rest of the paper is organized as follows: \Cref{sec:background} introduces the key concepts required to understand our proposal. \Cref{sec:related-work} reviews the state-of-the-art, and \Cref{sec:methodology} details our approach. \Cref{sec:method2,sec:results} present the experimental setup and results, while \Cref{sec:conclusions} discusses limitations and future work.

\section{Background}
\label{sec:background}

This section presents the necessary background for understanding the proposed work (notation summary is available in \Cref{tab:notation_summary}). Since radio link behavior and external environmental factors evolve over time, RLF prediction operates on time series data. 

\begin{table}[h]
\centering
\caption{Summary of notation.}
\label{tab:notation_summary}
\begin{tabular}{ll}
\toprule
\textbf{Notation} & \\ \midrule
$\mathcal{X}, \mathbf{X}, \mathbf{x}, x$ & time series dataset, instance, signal, entry \\
$N, C, T$ & number of instances, channels, timestamps \\
$i, c, t$ & indices for time series, channels, timestamps \\
$\mathbf{y}, y$ & ground truths vector, instance \\
$\hat{\mathbf{y}}, \hat{y}$ & predicted failure vector, instance \\
$\mathbf{\Phi}, \phi$ & saliency map, value \\
$\psi_c,\psi_{n,c}$ & global, local channel importance \\
$\mathbf{I}, \mathbf{S}$ & set of all feature indices, feature subset \\
$P$ & number of Monte Carlo samples \\
$\mathbf{X}_{\mathbf{S}}, \tilde{\mathbf{X}}$ & masked input, pruned feature set \\
$\tau$ & pruning threshold \\
$M$ & flattened feature dimension \\
$u, v, \mathcal{N}_K(v)$ & node, neighbor, set of $K$ nearest neighbors \\
$\mathbf{h}_v$ & aggregated embedding for node $v$ \\
\bottomrule
\end{tabular}
\end{table}

\begin{definition}[Time Series Data]
A time series dataset, $\mathcal{X}=
\{\mathbf{X}_1, \ldots, \mathbf{X}_N\} \in \mathbb{R}^{N \times C \times T},$ is a collection of $N$ time series.
A time series, $\mathbf{X}$, is a collection of $C$ signals (or channels), $ \mathbf{X}=
\{\mathbf{x}_1, \ldots, \mathbf{x}_C\} \in \mathbb{R}^{C \times T}$. Each signal, $\mathbf{x}$, is a sequence of $T$ observations,
$\mathbf{x}=[x_1, \ldots, x_T] \in \mathbb{R}^T$.
\end{definition}

If $C > 1$ the time series is termed multivariate.
Time series datasets have broad applications, including forecasting, anomaly detection, and classification. In this study, we frame next-day RLF prediction as a binary classification task, where the goal is to determine whether a given radio link will experience a failure within the next 24 hours, based on historical observations.


\begin{definition}[RLF Prediction]
Let $\mathcal{X}$ be a multivariate time series dataset. The goal of RLF prediction is to learn a function $f: \mathcal{X} \rightarrow [0,1]$ that assigns to each input $\mathbf{X}_n$ a probability of failure within the next day $\mathbf{\hat{y}} = [f(\mathbf{X}_1), \ldots, f(\mathbf{X}_N)] \in [0,1]^N$, as close as possible to the real status, $\mathbf{y}\in \{0,1\}^N$, where ${y} = 1$ indicates failure and ${y} = 0$ indicates normal operation. 
\end{definition}

The RLF prediction task can be addressed with various time series classification models, depending on dataset complexity. Recent work by Hasan \textit{et al.} \cite{hasan2023transformer} demonstrates that Transformer-based models are particularly effective for this task. Their multi-head attention mechanism captures long-range temporal dependencies by contextualizing each input element with respect to past and future observations, while multiple attention heads enable learning from diverse representation subspaces to identify complex temporal patterns \cite{zerveas2021transformer}. While very effective, neural-network based predictions remain opaque from the human perspective, whereas peering inside their decision making could be beneficial for understanding the RLF causes.
In this sense, there has been increasing interest in finding ways for explaining time series classifiers through post-hoc, model-agnostic approaches. The most common approach is through saliency maps, which provide feature for each time series observation~\cite{theissler2022explainable}.

\begin{definition}[Saliency Map]
Given a time series $\mathbf{X} \in \mathbb{R}^{C \times T}$, a saliency map is a matrix $\mathbf{\Phi} \in \mathbb{R}^{C \times T}$, where each entry $\phi_{c,t} \in \mathbf{\Phi}$ represents the relevance of the observation for channel $c$, at timestamp $t$, i.e., $x_{c,t}$ , with respect to the RLF prediction.
\end{definition}

One of the most common ways of estimating such relevance scores is using SHAP \cite{lundberg2017unified}, which generates feature importance as the expected marginal contribution that is the weighted average of the feature's contribution for a given instance. Formally:
\begin{equation}
\label{eq:shap}
\phi_{c,t} = \sum_{\mathbf{S} \subseteq \mathbf{I} \setminus \{(c,t)\}} \! \frac{(|\mathbf{I}| - |\mathbf{S}| - 1)! \, |\mathbf{S}|!}{|\mathbf{I}|!} \bigl(f(\mathbf{X}_{\mathbf{S} \cup \{(c,t)\}}) - f(\mathbf{X}_\mathbf{S})\bigr)
\end{equation}
\noindent where, $\phi_{c,t}$ denotes the SHAP value for the observation indexed by $(c,t)$, which quantifies its contribution to the model prediction. The set $\mathbf{I}$ represents the collection of all feature indices, with cardinality $|\mathbf{I}| = C \cdot T$. The summation is taken for all subsets $\mathbf{S}$ of $\mathbf{I}$ that exclude the feature $(c,t)$. The term $f(\mathbf{X}_\mathbf{S})$ refers to the model's prediction when only the features in subset $\mathbf{S}$ are used (with the remaining features typically masked or replaced by baseline values), and the difference $f(\mathbf{X}_{\mathbf{S} \cup {(c,t)}}) - f(\mathbf{X}_\mathbf{S})$ represents the marginal contribution of the feature $(c,t)$.

SHAP values are grounded in an additive explanation model, where the original model output is approximated by a sum of feature contributions. That is,
$f(\mathbf{X}) \approx \phi_0 + \sum_{(c,t) \in \mathbf{I}} \phi_{c,t}$, where $f(\mathbf{X})$ is the model output for the complete input $\mathbf{X}$, $\phi_{c,t}$ is the attribution assigned to feature $(c,t)$, and $\phi_0$ is the expected model output over the background dataset (i.e., the baseline prediction in the absence of any features).
Since the exact computation of SHAP values is combinatorially unfeasible, practical implementations rely on approximation techniques. 
In particular, the Sampling Explainer~\cite{lundberg2017unified} approximates SHAP values via Monte Carlo sampling of feature coalitions. Assuming feature independence, this method enables efficient estimation even when using large, high-dimensional, background datasets. 


\section{Related Work}
\label{sec:related-work}

Related work falls into two main areas: predictive models for failure detection in 5G radio access networks (RAN) and explainable AI (XAI) methods for interpreting anomaly and failure predictions.

Within the first area, several studies have explored machine learning-based RLF prediction in 5G RAN. Early work by Boutiba \textit{et al.} \cite{boutiba2021radio} combined LSTM and SVMs to model correlations between weather data and radio link measurements but did not distinguish between time series and categorical inputs, limiting temporal modeling. Aktacs \textit{et al.} \cite{aktacs2022towards} addressed this by preprocessing weather and radio link data separately and training a deep LSTM+ model on ten days of historical data. Islam \textit{et al.} \cite{islam2022deep} further improved performance with an LSTM autoencoder using five days of input, showing that shorter temporal windows can be more effective.
Despite their improvements, LSTM-based models face vanishing gradient issues, lack explicit feature importance, and offer limited interpretability. To address these challenges, Hasan \textit{et al.} \cite{hasan2023transformer} proposed GenTrap, a hybrid GNN–Transformer architecture that combines spatial aggregation from radio links and weather stations with temporal pattern extraction from multivariate time series. Although GenTrap outperforms LSTM-based methods, it does not provide interpretability regarding its decision process.

Explainability is critical in the prediction of failures, particularly in network operations, where understanding the decisions of the model is essential for trust and intervention. Classical methods for anomaly detection are often criticized for lack of transparency~\cite{zou2020explainable}. Thus, Zou et al.~\cite{zou2020explainable} applied SHAP values to explain the output of decision trees and LSTM-based DeepLog models, extracting meaningful event sequences and feature contributions. However, their use of classical SHAP limits applicability to complex multivariate time series traffic.
Roshan et al.~\cite{roshan2022using} used Kernel SHAP to identify key features driving unknown attack detection, improving both performance and interpretability. Antwarg et al.~\cite{antwarg2021explaining} applied SHAP to an autoencoder-based anomaly detection, proposing an architecture that attributes reconstruction error to specific features and produces interpretable anomaly scores. In the industrial domain, SHAP has also been used to enhance failure detection in prognostic and health management systems~\cite{nor2022abnormality}. However, these studies focus on univariate or shallow models and do not generalize well to deep architectures that handle multivariate time series. 
More recent work has begun exploring explainability in mobile network traffic forecasting. AICHRONOLENS~\cite{fiandrino2024aichronolens} and DeExp~\cite{gholian2025deexp} introduced explainable models for anomaly-aware forecasting in mobile traffic, while Marantis et al.~\cite{marantis2024ran} combined LSTM, Isolation Forest, and SHAP to detect traffic anomalies in 6G networks. However, these models are limited to univariate input and do not scale to the complexity of failure prediction in 5G RAN. TimeSHAP \cite{bento2021timeshap} and TSHAP \cite{nguyen2025tshap} focus on multivariate time series and provide an approximation of important time steps in long-time series without considering the state-of-the-art Transformer model to predict failures. On the other hand,
CHRONOPROF \cite{perez2025chronoprof} is a generalized explainable architecture to visualize model behaviors for univariate and multivariate traffic forecasting for resource allocation and management; however, it neither quantifies explanation reliability nor suggests explanations to refine or simplify models.

\section{Workflow and Methodology}
\label{sec:methodology}

\begin{figure*} [h]
    \centering
    \includegraphics[width=.9\textwidth]{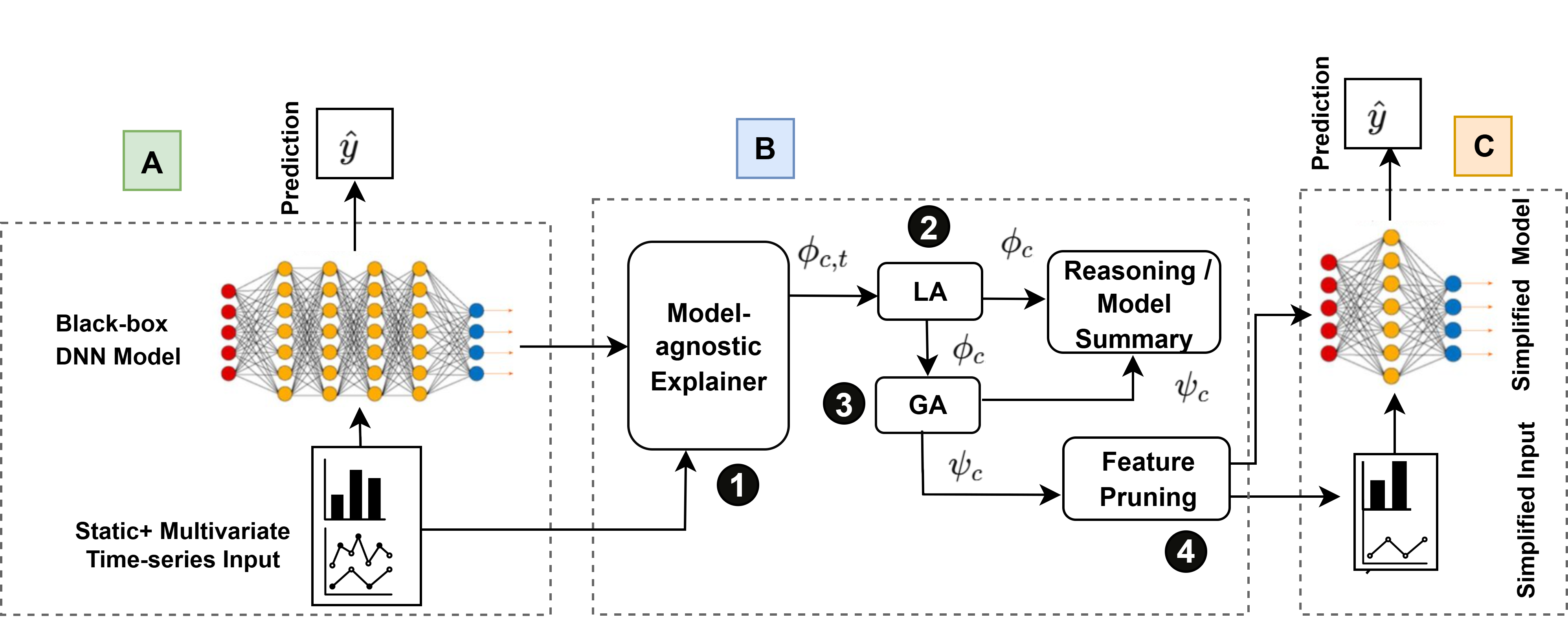}
    \caption{\framework architecture with three components: a black-box DNN, an explainer, and an explainability-guided simplified DNN. The explainer includes four modules: model-agnostic explainer, Local Aggregation (LA), Global Aggregation (GA), and feature pruning. The output is a simplified DNN with reduced components.}
      \label{fig:workflow}
\end{figure*}

In this section, we present a high-level overview of \framework. The main architecture is shown in Fig.~\ref{fig:workflow} and consists of three components: a black-box DNN model, an explainer, and an explainability-guided simplified DNN model. 
\framework can support any DNN like Transformer, GNN, and LSTM with any saliency-based model-agnostic explainer. 
The explainer is the core of \framework with four modules: \circled{1} a model-agnostic explainer, \circled{2} a \textit{Local Aggregation (LA)} module, \circled{3} a \textit{Global Aggregation (GA)} module, and \circled{4} a feature pruning module.
The explainer module uses any model-agnostic explainer convenient for time series and large volumes of data, to compute feature relevance score at each time step. LA performs necessary local feature importance and reasoning, whereas GA performs necessary aggregation to provide global feature importance and model summary. The last module recommends feature-pruning based on the global scores to remove unnecessary model components. In the following, we elaborate on the operation of each component of the explainer while integrated with a given failure prediction model.

\subsection{Model-agnostic Explainer Module} 

The explainer module takes two inputs: a deep neural network-based prediction model and its multivariate input. Note that in the RLF prediction, the multivariate data consists of static categorical (spatial) and time-series (temporal) features. The feature importance for the given model and input can be processed using any SHAP-based explainer, but we choose model-agnostic Sampling Explainer (SE) \cite{lundberg2017unified} due to its flexibility of applying on any model and efficiency for processing large data set. 
In particular, given a time series $\mathbf{X}$, the output of this module is a local explanation ${\Phi}$, containing the contribution of each observation in the time series towards a failure outcome. The positive SHAP values indicate a contribution toward a failure and negative ones indicate a contribution toward normal operation while a value around $0$ indicates that the feature is not relevant.  

SE computes such SHAP values by approximating \Cref{eq:shap} using Monte Carlo approximation: ${\phi}_{c,t} = \frac{1}{P} \sum_{p=1}^{P} \left[ f\left(\mathbf{X}_{\mathbf{S}_p \cup \{(c,t)\}}\right) - f\left(\mathbf{X}_{\mathbf{S}_p}\right) \right]$. Here, $P$ denotes the number of random permutation samples, and each $\mathbf{S}_p \subseteq \mathbf{I} \setminus \{(c,t)\}$ is a subset of the feature index set that excludes the feature $(c,t)$ under evaluation. The quantity ${\phi}_{c,t}$ estimates the contribution of feature $(c,t)$ to failure by averaging its marginal impact across $P$ random subsets of the remaining features. 
Since features cannot be literally removed from an input instance, we simulate their exclusion using background data, which serves as a reference distribution. Specifically, for features not included in $\mathbf{S}_p$, their values are replaced with corresponding values drawn from the background distribution. This allows us to isolate the effect of including $(c,t)$ in each subset $\mathbf{S}_p$ and quantify its influence on the model prediction $f(\cdot)$. 



\subsection{Local Aggregation (LA) and Global Aggregation (GA)}

A saliency map $\mathbf{\Phi}$ represents the importance of input features for a given prediction by calculating their contributions over $T$ time steps for $C$ features, resulting in a map of shape $C \times T$. This allows both local and global attribution of feature relevance.

\begin{equation}\label{local}
\psi_{n,c}
=
\sum_{t=1}^{T}
\left(
(1-\alpha)\,\phi_{n,c,t}
+
\alpha\,\lvert \phi_{n,c,t} \rvert
\right), \alpha \in \{0,1\}
\end{equation}
\begin{equation}\label{global}
\psi_c = \frac{1}{N}\sum_{n=1}^{N}\psi_{n,c}
\end{equation}

The first term $\psi_{n,c}$ corresponds to local aggregation (\circled{2}), which captures the aggregated contribution of each feature $c$ across all time steps for a single prediction $i$. This local aggregation either uses absolute or standard summation, based on the problem requirement, where $\alpha = 1 $ means absolute summation, and $\alpha = 0 $ means standard summation. This provides a local explanation, i.e., an attribution of feature importance specific to an individual prediction, which can be valuable for case-level analysis. However, such explanations are limited in capturing the model’s overall behavior, especially in time series data where features span multiple time steps.

The second term $\psi_c$ corresponds to global aggregation (\circled{3}), obtained by averaging the local contributions of each feature across $N$ samples. This yields a global measure of feature importance that highlights which features consistently influence the model’s predictions, enabling a broader and more interpretable understanding of model behavior.

\subsection{Feature Pruning}

Explanations generally serve two complementary purposes. First, local explanations provide reasoning behind individual predictions, enabling users to understand why a specific decision was made. Second, global explanations offer insights into the model’s overall behavior, in particular by identifying the features that exert the greatest influence on outcomes.

Building on this perspective, we propose a feature selection strategy that exploits global feature importance scores $\boldsymbol\psi$ to derive a more compact and efficient model representation. Specifically, Feature Pruning \circled{4} introduces an explanation-aware scheme for eliminating redundant or low-utility features. Since not all features contribute meaningfully to failure prediction, retaining irrelevant ones may unnecessarily increase model complexity and even degrade predictive performance. To address this, we remove features whose global importance score $\psi_c$ falls below a user-defined or adaptively learned threshold. Formally, we construct a refined feature set consisting only of the most relevant features: $\tilde{\mathbf{X}} = \{\mathbf{x}_c\mid |\psi_c| > \tau \}$. 
Here, $\tau$ is a user-defined threshold to prune unnecessary features. $\tilde{\mathbf{X}}$ is the new feature set with important features that we obtain after pruning.

After pruning unnecessary features, the prediction models can themselves be simplified by identifying and discarding architectural components dedicated to those pruned features. For instance, if half of the time-series features are deemed irrelevant, we can reduce the dimensionality of the embedding vectors in transformer models or decrease the number of encoder layers in recurrent models used to represent each input token. Such reductions naturally lower the number of hyperparameters and lead to a more streamlined architecture. In this way, we revisit and adapt existing models by removing redundant processing steps or trimming neural layers, thereby generating more efficient models without compromising predictive quality.

\section{Informed simplification of DNNs}
\label{sec:method2}

In this section we apply \framework on two commonly used, state-of-the-art DNNs for 5G-RAN data, to illustrate how, together with domain-expert knowledge, \framework can simplify complex black-boxes to realize explainability-guided lighter models with significantly fewer hyperparameters. First, we present the experimental details and then outline the architecture and operation of these models, followed by their integration into \framework to have the final explainability-guided prediction models with better performance at scale. We implement all models and explainers using Tensorflow and SHAP libraries\footnote{Code is available at \url{https://github.com/PINetDalhousie/prometheus.git}}. The models are deployed and tested on an Intel(R) Xeon(R) Silver 4310 CPU @ 2.10GHz with OS Ubuntu 24.04.2 LTS and a server with a NVIDIA-SMI 560.35.05 GPU and CUDA Version 12.6. Numerical results are presented and discussed in \Cref{sec:results}.

\textbf{Dataset.} We adopt the Turkcell open source dataset, a major telecommunication provider, as described in \cite{aktacs2022towards, islam2022deep, hasan2023transformer}. This dataset encompasses two deployments, urban and rural, containing observations from radio and weather stations. The urban deployment includes 1674 base stations and 20 surrounding weather stations for a total of 1.8 million data points, while the rural deployment consists of 1388 base stations and 117 weather stations, for a total of 0.4 million observations.
The dataset includes six tables with radio link (RL) and weather station (WS) features. RL tables provide static site parameters and time-series KPIs, while WS tables contain station metadata, hourly observations (aggregated daily), forecasts, and inter-station distances. After preprocessing, we obtain 17 time series features supplemented with categorical variables. Specifically, there are 7 radio link KPI time series, 9 weather station time series features, including temperature, precipitation, humidity, and wind speed, and a positional encoding feature for time-step alignment in the transformer model. The complete list of RL KPI features used in the prediction models is provided in Table \ref{tab:rl}.
We use both deployments (urban and rural) for testing different models and evaluating the performance of the models using precision, recall, and F1-score. 

\begin{table}[t]
    \centering
      \begin{tabular}{l|p{0.6\columnwidth}} \toprule
\textbf{kpis} & \textbf{description}    \\ \midrule
severely Error\_second &  Count of 1 sec periods with high density error  \\
error\_second & Count of 1 sec periods with error \\
unavail\_second & RL unavailable operation duration in seconds\\
bbe & Indicator of performance degradation, Background bit error count\\
rxlevmax & RL received power level\\
capacity & Data throughput capability of an RL\\
\bottomrule
\end{tabular}
\caption{The list of radio links KPIs. }
\label{tab:rl}
\end{table}

\textbf{Model Setup.} In the training phase, the datasets are divided into five folds using the rolling origin technique \cite{hasan2023transformer}. The first fold (F4) uses the initial 70\% of the data for training, the following 20\% for validation, and the final 10\% for testing. The remaining folds (F3, F2, F1, and F0) are constructed by progressively discarding the last 10\% of data from the preceding fold. Rolling origin is particularly suitable for time-series prediction as it preserves the temporal order of the data while allowing the model to be evaluated on multiple sequential splits. Each model is trained across all folds for every deployment to ensure robustness and consistency.
We train each model with a batch size of 1024, varying the number of epochs from 1000 to 4000, and keeping other parameters fixed following existing prediction solutions \cite{islam2022deep, hasan2023transformer}. The optimal hyperparameter list is presented in Table~\ref{tb:para}. Once a model has been trained, we first assessed the impact of the number of previous days on the prediction, in Table \ref{tb:bestday}. The result suggests that the previous four days' historical context offers the best prediction; thus, we used this parameter in the rest of the evaluations. 


\begin{table}[h]
    \centering
    \begin{tabular}{ll}
        \toprule
        \textbf{Parameters} & \textbf{Values} \\
        \midrule
        Batch size & 1024 \\
        Learning rate, $\sigma$ & 0.001 \\
        Epochs & 1000--4000 \\
        Number of weather stations $(k)$ & 3 \\
        Number of previous days $(n)$ & 4 \\
        \bottomrule
    \end{tabular}
    \caption{The list of hyperparameters in experiments.}
    \label{tb:para}
\end{table}


\begin{table}[h]
    \centering
    \begin{tabular}{cccc}
        \toprule
        & \multicolumn{3}{c}{\textbf{Rural Data Set}} \\
        \midrule
        \textbf{Previous days} & \textbf{Precision} & \textbf{Recall} & \textbf{F1 Score} \\
        \midrule
         \rowcolor{gray!30}
        4 & 0.88 $\pm$ 0.06 & 0.92 $\pm$ 0.07 & 0.90 $\pm$ 0.03 \\
        5 & 0.88 $\pm$ 0.05 & 0.88 $\pm$ 0.07 & 0.88 $\pm$ 0.03 \\
        6 & 0.86 $\pm$ 0.07 & 0.89 $\pm$ 0.07 & 0.87 $\pm$ 0.04 \\
        \bottomrule
    \end{tabular}
    \caption{Impact of the number of previous days on predictive performance in the Gentrap model.}
    \label{tb:bestday}
\end{table}

\textbf{Explainer Setup.}
For the explainer module, we adopt the SHAP SamplingExplainer~\cite{lundberg2017unified}. This module computes saliency maps for all failure predictions. The Explainer is fitted using the trained RLF model and the background dataset. Since failure and non-failure samples are highly imbalanced (0.3\% failure for the rural and 0.06\% failure for the urban dataset only), we focus on \textit{True Positive (TP)} predictions (correctly identified failures). Saliency maps are generated for these \textit{TP} cases in each fold’s test set to quantify feature importance. We use standard summation ($\alpha=0$) to generate the map in equation \ref{local} as we consider only \textit{TP} cases. By aggregating these maps using equation \ref{global}, we derive global feature importance scores. For feature pruning, we define $\tau$ such that it retains features that explain 95\% of total importance by using the cumulative sum of importance.

\subsection{Transformer-based Model}

\begin{figure}[ht]
    \centering
    \begin{subfigure}[t]{0.48\textwidth}
        \centering
        \includegraphics[width=.85\textwidth]{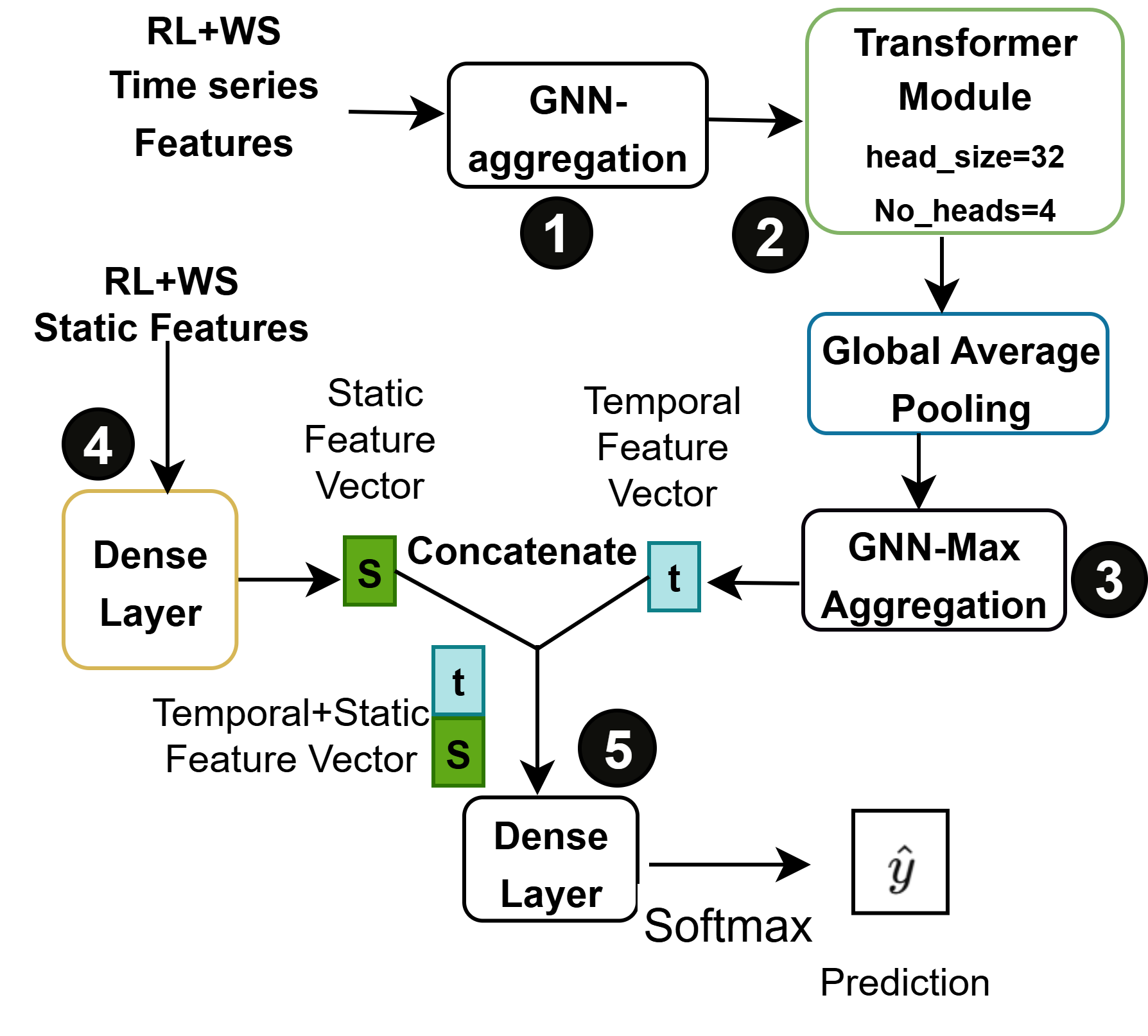}
        \caption{GenTrap.}
        \label{fig:architecture}
    \end{subfigure}
    \hfill
    \begin{subfigure}[t]{0.48\textwidth}
        \centering
        \includegraphics[width=0.6\textwidth]{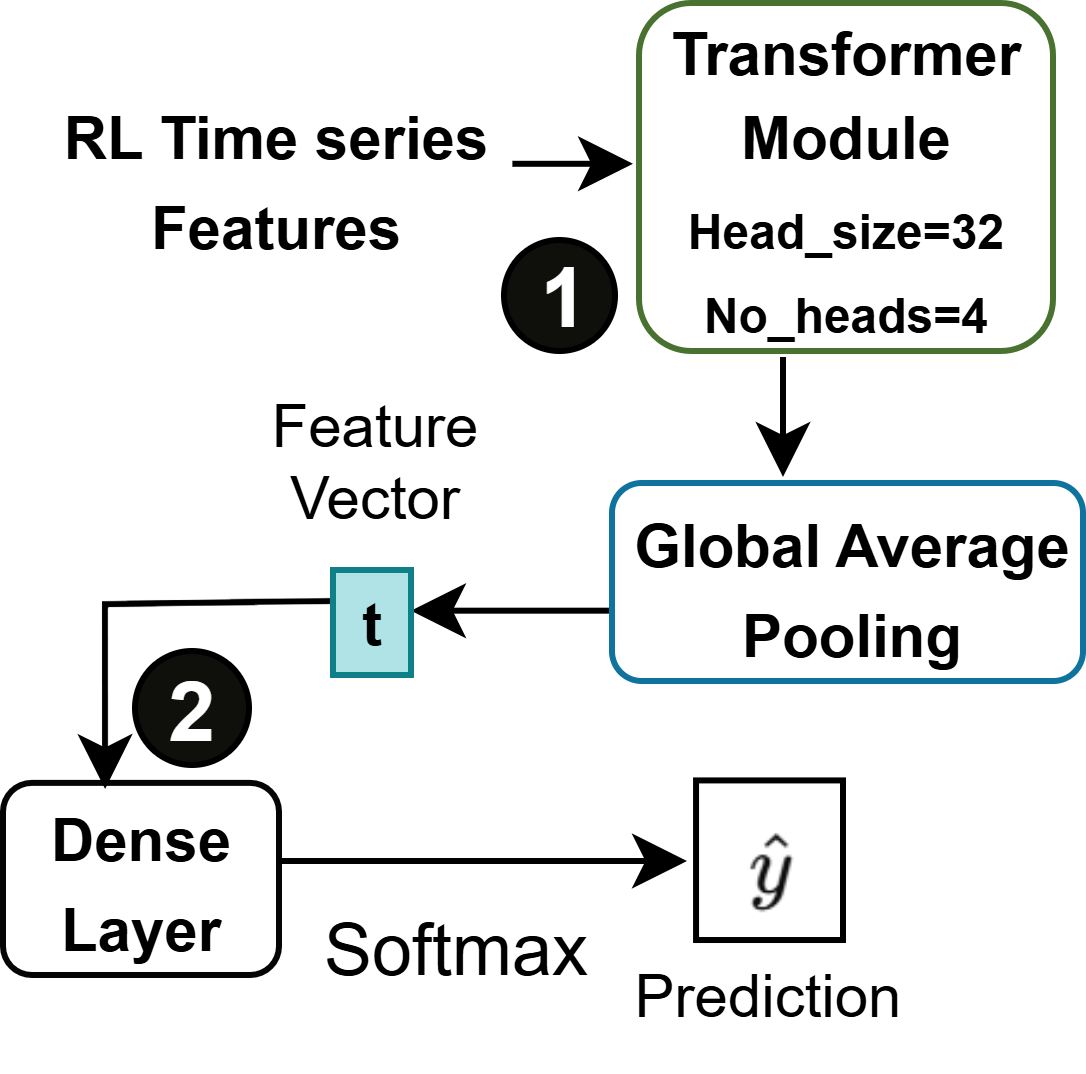}
        \caption{LTrans.}
        \label{fig:newarchitecture}
    \end{subfigure}
    \caption{Transformer-based architectures before and after \framework simplification.}
    \label{fig:architectures_comparison}
\end{figure}

Fig.~\ref{fig:architecture} presents a GNN-Transformer based state-of-the-art RLF prediction architecture, GenTrap \cite{hasan2023transformer}, consisting of four modules: \circled{1} GNN-Aggregation for closest weather station data capture for a RL, \circled{2} Transformer and \circled{3} GNN-max aggregation for time series feature representation, \circled{4} Dense layer for static feature vector, and \circled{5} another dense layer to calculate the final feature vector.

\textbf{GNN-Aggregation.}
The first stage of the GenTrap pipeline is the GNN-Aggregation module \circled{1} that takes as input a multivariate time series dataset of shape $N \times T$. The module integrates RL and WS features using a Graph Neural Network (GNN) with max aggregation. The GNN operates on a graph where each node corresponds to a weather station and edges encode spatial proximity. For each node $v$, the representation is computed by aggregating features from its neighboring weather stations: $\mathbf{h}_v = \max_{u \in \mathcal{N}_K(v)} \mathbf{h}_u$ with $\mathbf{h}_u = \sigma\left( \mathbf{W} \mathbf{X}_u \right)$. 
Here, $\mathcal{N}_K(v)$ denotes the set of $K$-neighboring weather stations closest to radio link site $v$, $\mathbf{X}_u$ contains the signals coming from weather station $u$, $\mathbf{W}$ is a learnable weight matrix, and $\sigma$ is a non-linear activation. The element-wise max operator aggregates the transformed neighbor features into a unified embedding $\mathbf{h}_v$.
The resulting input tensor for downstream processing has shape $N \times (R \cdot K) \times T$, i.e., $N$ time series, comprising $C=R \cdot K$ channels (radio links and weather station features), and $T$ timestamps.

\textbf{GNN-Transformer module.}
The Transformer module \circled{2} generates context-aware representations of the input sequence, which are then passed through a global average pooling layer to capture temporal dependencies for each (WS+RL) pair. To further aggregate these temporal embeddings, GNN-Max pooling \circled{3} is applied across the $K$ embedding vectors corresponding to the (WS+RL) pairs, producing a single compact feature vector. In parallel, the static features of radio links and weather stations are processed through dense layers \circled{4}, generating a static feature representation. The outputs of the temporal branch (Transformer) and the static branch are then concatenated to jointly capture both static and temporal dependencies. This fused vector is passed through an additional dense layer \circled{5}, followed by a Softmax layer that produces the final probability distribution for link failure prediction \cite{hasan2023transformer}.
This prediction model requires a total of 13.4K trainable parameters; thus, we will integrate it to \framework to assess the feature importance for a potential reduction in the feature space and the required parameters. 

\textbf{Explaining GenTrap.}
We begin by flattening the input tensor of shape $N \times (R \cdot K) \times T$ into a general form of $N \times M$, where $M = C \times T$ and $C = (R \cdot K)$. A wrapper function is then applied to ensure compatibility between the model’s original shape and the explainer, allowing the model output to align with the explainer’s requirements. Using this setup, we provide both the background data and the trained model (via the wrapper) to train the explainer model.
For local feature importance, a single input is passed to the explainer model at a time. The explainer computes the importance scores according to the Monte-Carlo approximation of \Cref{eq:shap}, producing a feature importance vector of shape $1 \times M$. This vector is reshaped to $C \times T$ to form a saliency map, where each value ${\phi_{c,t}}$ corresponds to the contribution of feature $c$ at time step $t$. Repeating this process for $N$ samples yields $N \times C \times T$ feature importance vectors (i.e., $N$ saliency maps), which are then aggregated to derive the global feature importance.
From the global feature importance, we identify the most relevant features for prediction and prune those that contribute very little or nearly zero. For instance, the analysis revealed that WS data and static features have minimal impact on prediction outcomes. Consequently, we prune these features and revisit the model architecture to eliminate redundant layers or steps associated with them. This process leads to a more compact design, motivating the development of a simplified transformer-only model tailored for the prediction task.

\textbf{LTrans.}
Figure \ref{fig:newarchitecture} illustrates the lightweight transformer-based RLF prediction model enhanced with XAI. While it preserves the two-module structure, each module is significantly simpler than in the original architecture. Since WS data is no longer used, there is no need for GNN-based aggregation to capture nearby WS features. Instead, the model operates solely on RL time-series data, resulting in a more compact input of shape $N \times C \times T$, where $C = R$ (radio link features only) and $T$ denotes the number of timestamps.
Module \circled{1} consists of a streamlined transformer branch for capturing temporal dependencies, followed by a dense layer in Module \circled{2}, which generates the final failure probability vector. By discarding unnecessary inputs and processing steps, the proposed model reduces both architectural complexity and parameter count requiring only 5.8K trainable parameters compared to 13.4K in the previous model. This makes the new design not only more efficient and easier to train but also highly effective as a lightweight yet explainability-guided predictor of link failures.

\subsection{LSTM-based Model}

\begin{figure}[ht]
    \centering
    \begin{subfigure}[b]{0.45\textwidth}
        \centering
        \includegraphics[width=0.9\textwidth]{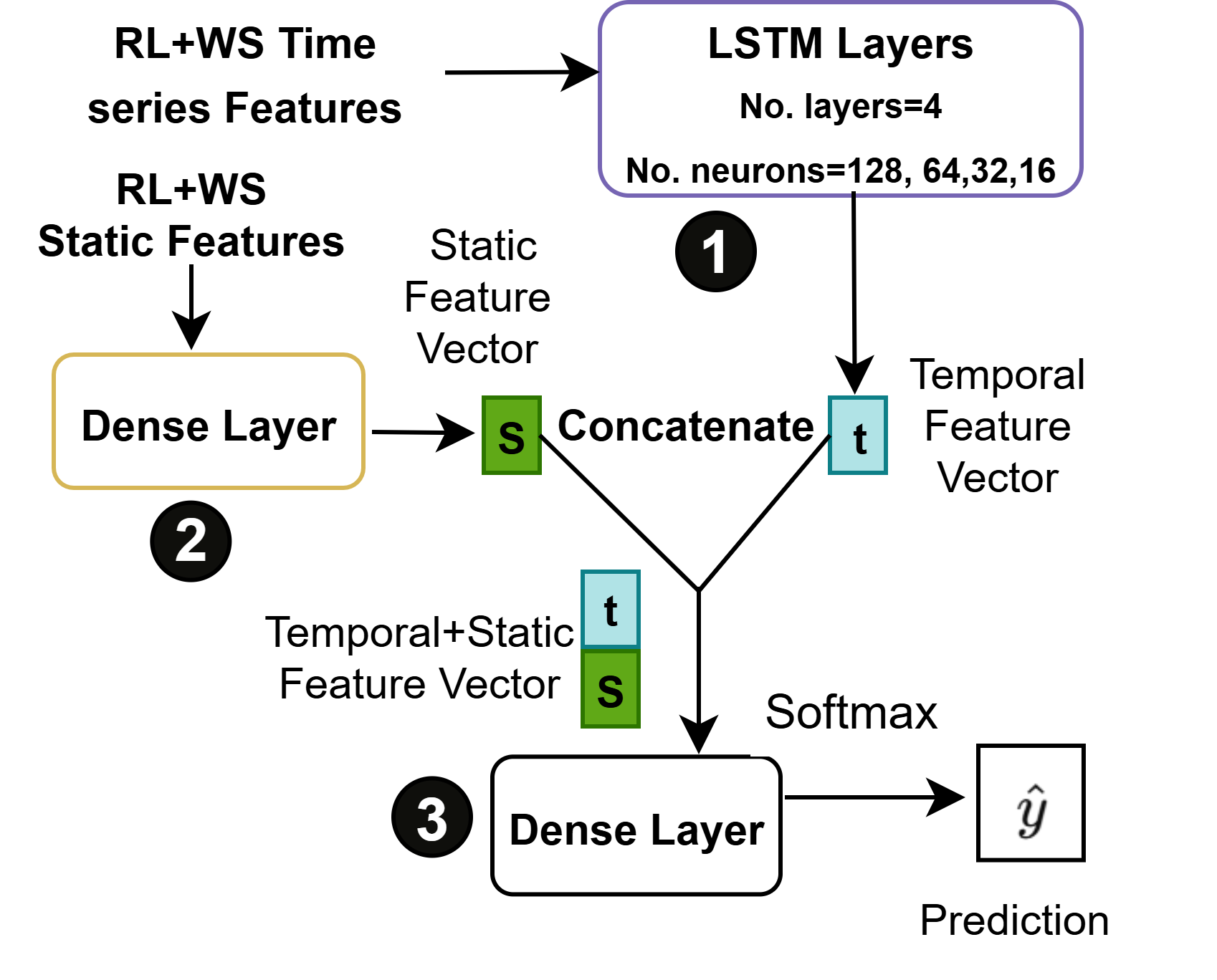}
        \caption{LSTM+.}
        \label{fig:lstm}
    \end{subfigure}
    \hfill
    \begin{subfigure}[b]{0.45\textwidth}
        \centering
        \includegraphics[width=0.7\textwidth]{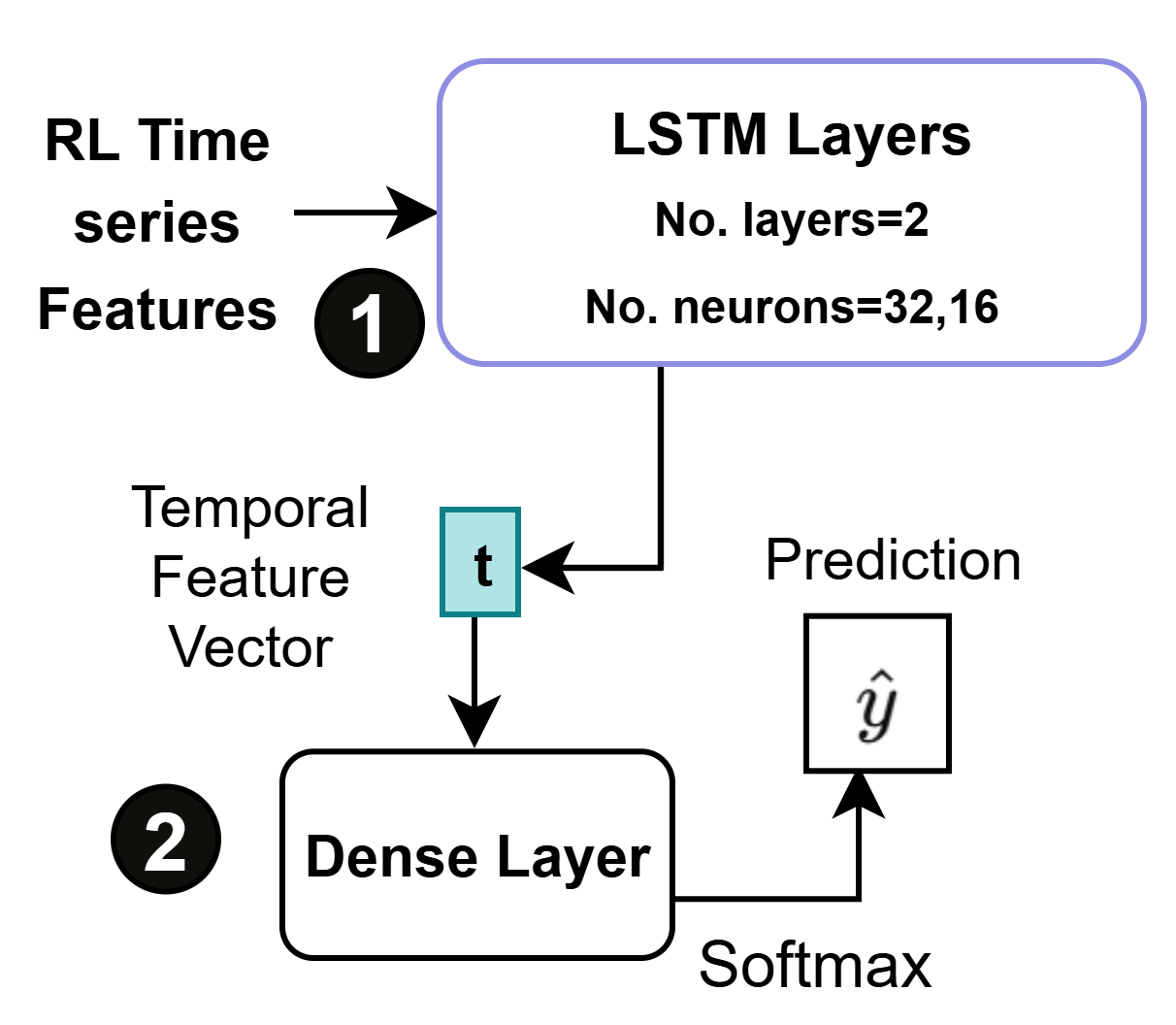}
        \caption{LLSTM+.}
        \label{fig:newlstm}
    \end{subfigure}
    \caption{LSTM-based architectures before and after simplification with \framework.}
    \label{fig:lstm_comparison}
\end{figure}

Figure \ref{fig:lstm} illustrates the traditional LSTM-based architecture, LSTM+, for the prediction of RLF \cite{aktacs2022towards}. The model is organized into three modules: (i) an LSTM module \circled{1} that processes the time-series features, (ii) a dense layer \circled{2} that incorporates static features, and (iii) a final dense layer \circled{3} that combines static and temporal representations to generate the failure probability.
The core of the model is the LSTM branch, which consists of four stacked encoder layers with 128, 64, 32, and 16 neurons, respectively, to extract temporal feature representations from the input sequence. Unlike the GNN-Transformer model, this architecture relies on derived WS features such as mean, minimum, maximum, and standard deviation from the closest weather stations. While effective, this design is computationally heavier: it requires approximately 154.7K trainable parameters to produce predictions, making it significantly more complex than both the GNN-Transformer and the lightweight Transformer-based alternatives.

\textbf{Explaining LSTM+.}
We integrate the LSTM-based (LSTM+) model into \framework to identify the most influential features for RLF prediction. The model input has a shape of $N \times C \times T$, which we flatten into a general input of size $N \times M$, where $M = C \times T$ and $C$ represents the set of all RL and derived WS features. Following the same procedure as before, we generate $N \times C \times T$ feature importance vectors, i.e., $N$ saliency maps across all samples, which are then aggregated to derive the global feature importance of the prediction model.
Using the global feature importance, we apply feature pruning similar to the GenTrap case. For the LSTM+ model, we observe that static features contribute minimally to the prediction outcome, whereas several time-series features play a more significant role. Consequently, we retain only the influential features and revisit the model architecture to remove layers and steps associated with pruned features. This refinement results in a more interpretable and lightweight LSTM model, which preserves predictive power while reducing complexity.

\textbf{LLSTM+.}
Figure \ref{fig:newlstm} illustrates the lightweight LSTM-based RLF prediction model derived from XAI. The architecture has been simplified to two modules, each significantly leaner than in the original model. Specifically, the model now leverages only 14 RL and WS time-series features. The resulting input has a shape of $N \times C \times T$, where $C$ corresponds to the selected RL and WS features and $T$ denotes the timestamps.
Module \circled{1} is the LSTM branch, now reduced to only two layers after pruning more than half of the time-series features. The dense layers previously used to capture static features are entirely removed. As a result, the streamlined model consists solely of the LSTM branch and a single dense layer \circled{2} to produce the failure probability vectors. This reduction not only simplifies the architecture but also accelerates training and inference. Overall, the lightweight model requires only 11K trainable parameters, representing nearly a 92\% reduction compared to the original LSTM+ model while maintaining predictive capability.

\section{Results and Discussion}
\label{sec:results}

This section presents the evaluation results by answering three research questions. \textbf{RQ1:} does RLF prediction require categorical features and weather context? \textbf{RQ2:} are lightweight models comparable to the black-box models in prediction accuracy? \textbf{RQ3:} which features contribute to the RLF prediction, and why?

\subsection{Does RLF prediction require categorical features and weather context? (RQ1)}

\begin{figure}[ht]
\centering
\begin{subfigure}[t]{0.405\textwidth}
    \centering
    \includegraphics[width=.9\textwidth]{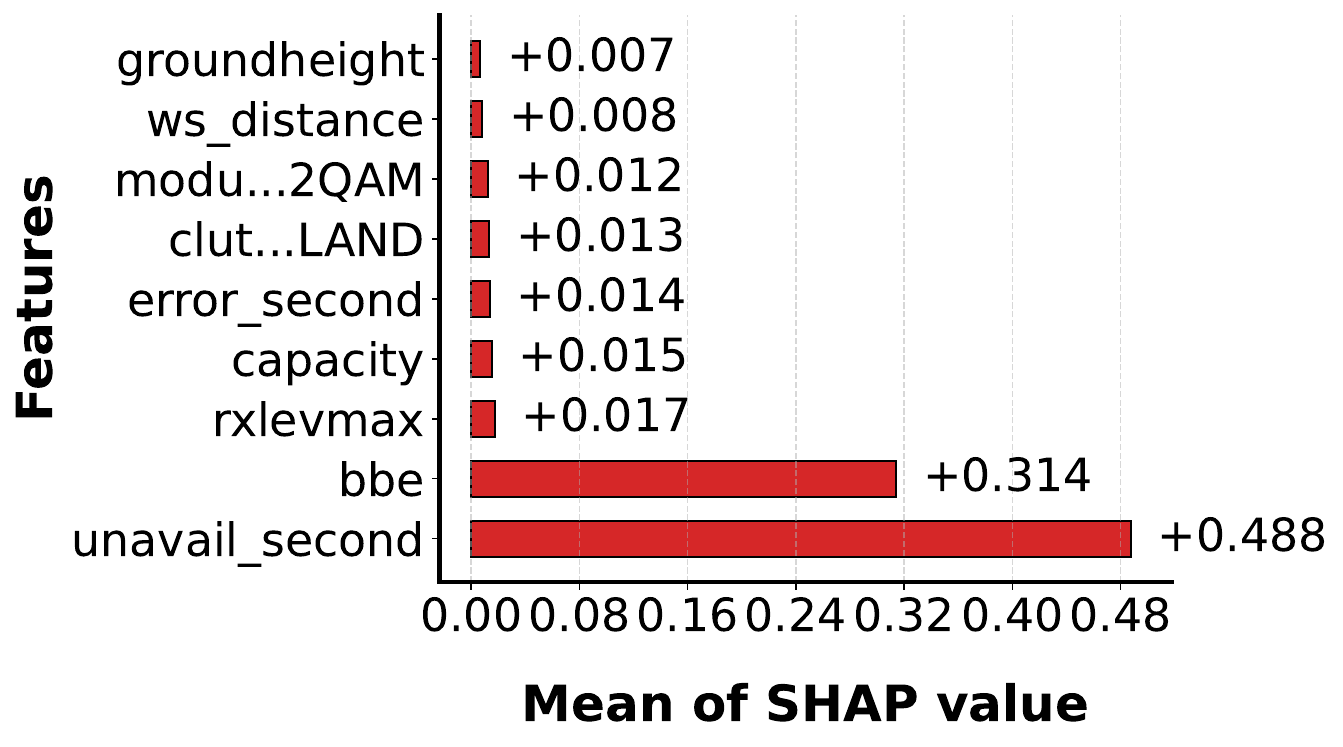}
    \caption{GenTrap}
    \label{fig:failure_sum}
\end{subfigure}
\hfill
\begin{subfigure}[t]{0.405\textwidth}
    \centering
    \includegraphics[width=.9\textwidth]{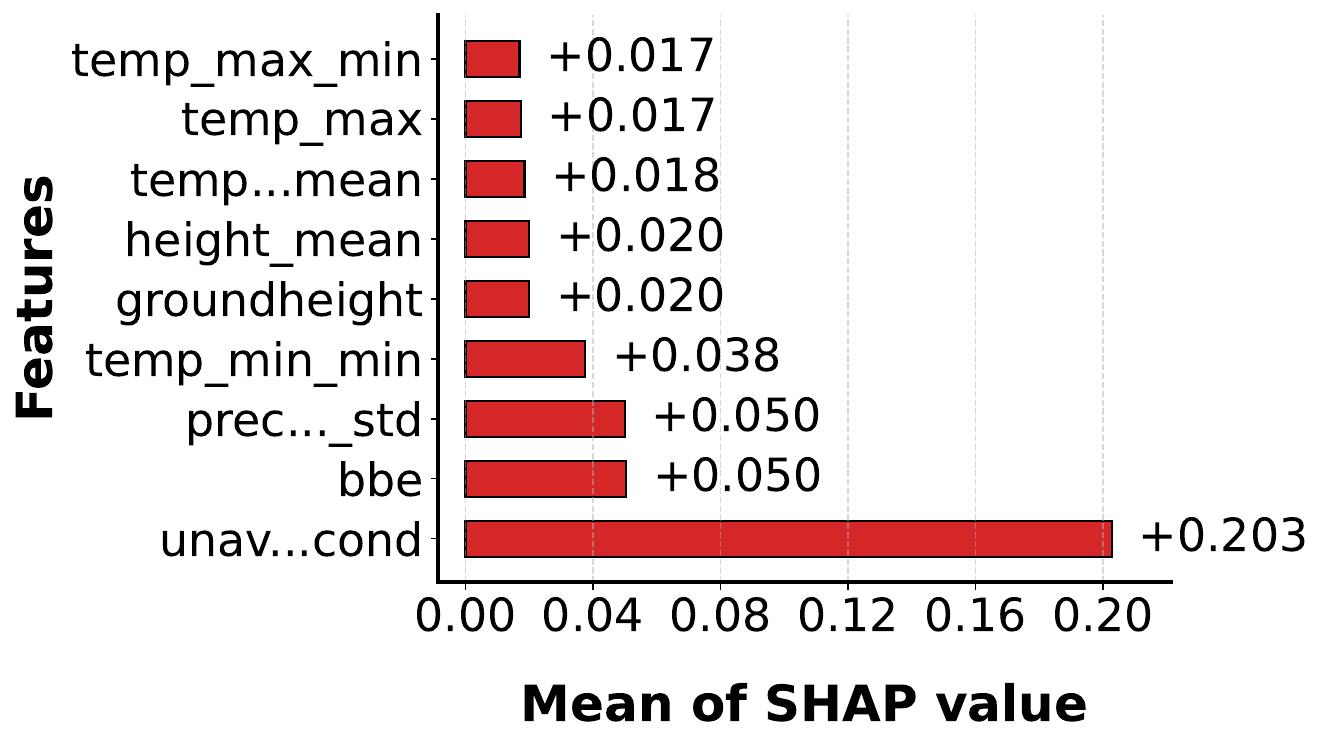}
    \caption{LSTM+}
    \label{fig:failure_sum_lstm}
\end{subfigure}
\caption{Average SHAP values across all failures for rural deployment.}
\label{fig:failure_sum_comparison}
\end{figure}

\begin{figure}[ht]
\centering
\begin{subfigure}[t]{0.405\textwidth}
    \centering
    \includegraphics[width=.9\textwidth]{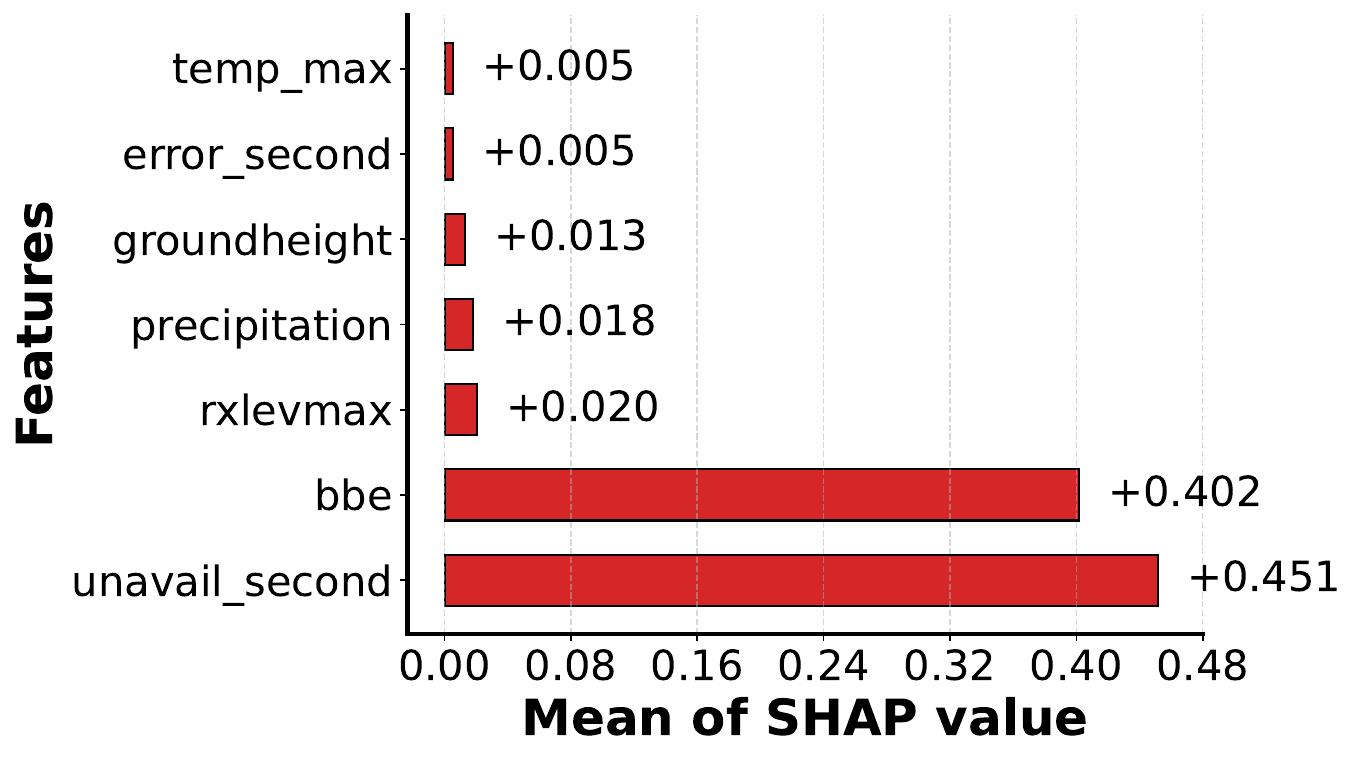}
    \caption{GenTrap}
    \label{fig:shap_gen_urban}
\end{subfigure}
\hfill
\begin{subfigure}[t]{0.405\textwidth}
    \centering
    \includegraphics[width=\textwidth]{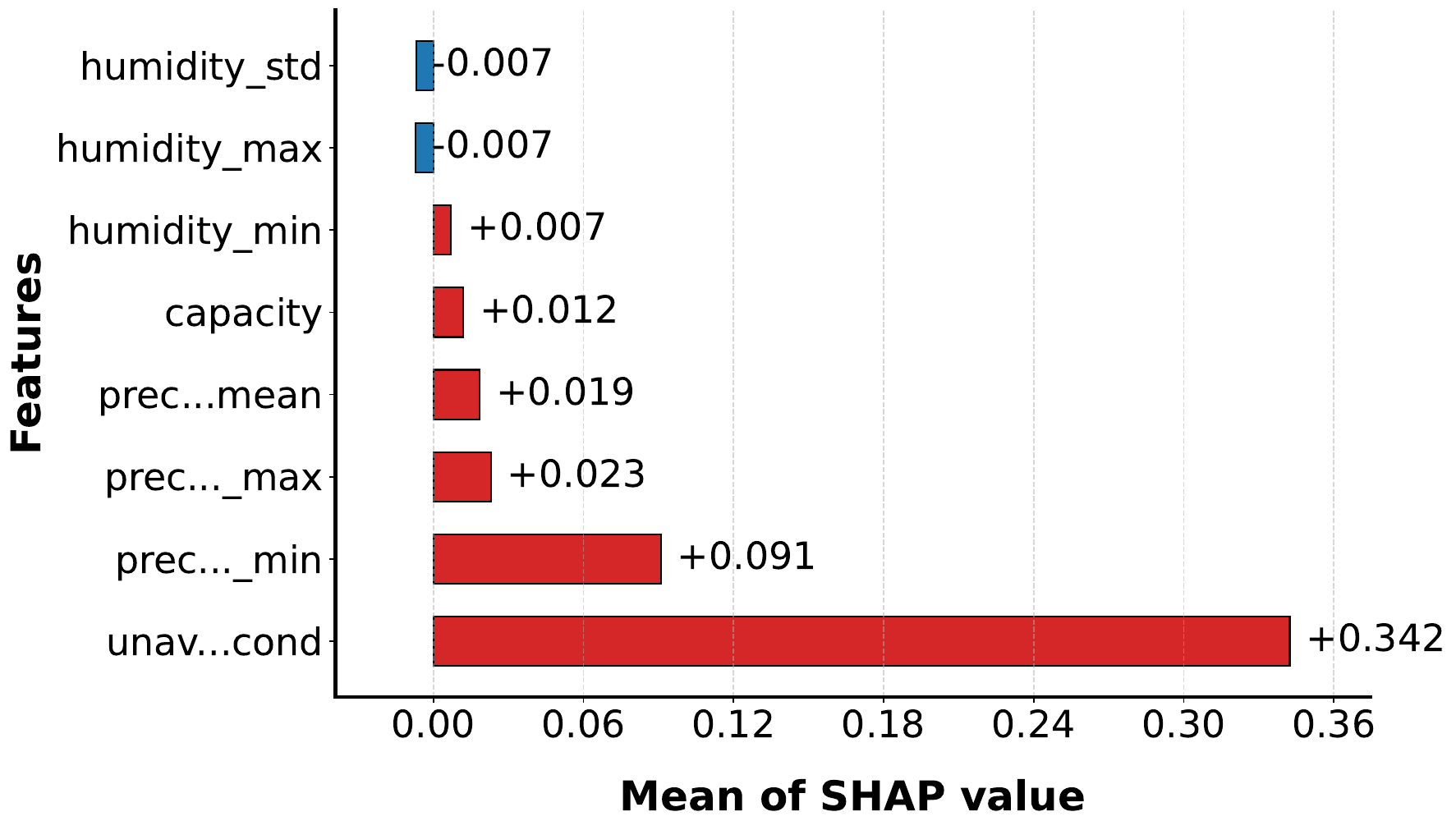}
    \caption{LSTM+}
    \label{fig:shap_lstm_urban}
\end{subfigure}
\caption{Average SHAP values across all failures for urban deployments.}
\label{fig:shap_urban}
\end{figure}

This section presents the global feature importance of GenTrap and LSTM+ to answer RQ1. Figure \ref{fig:failure_sum} and \ref{fig:shap_gen_urban} present the average SHAP values for the GenTrap (GNN-Transformer) model for rural and urban scenarios, respectively, where we keep the features specified by $\tau$. 
From the figure, we see that \texttt{unavail\_second} and \texttt{bbe} are the most important features for predicting failures with values +0.488 and +0.314, respectively (covering almost 86\% of the importance) respectively, along with some other influential features such as \texttt{capacity} or \texttt{rxlevmax}. On the other hand, WS features and static features have very little effect on the failure prediction, which have SHAP values close to zero, meaning that the model assigns them little to no importance for the prediction. Interestingly, we observe that the patterns of the influencing features that the model learned are mainly from RL KPIs, i.e., the changes in these KPIs contributing to the failures. The high impact of RL KPIs on failures is expected and supported by the existing literature \cite{ITUTG826, 3gpp-ts-36.331,feng2016millimetre}. However, we need a deeper investigation into this observation for an informative conclusion.

\begin{figure*}[h]
    \centering
    \begin{subfigure}[t]{0.24\linewidth}
        \centering
        \includegraphics[width=\linewidth]{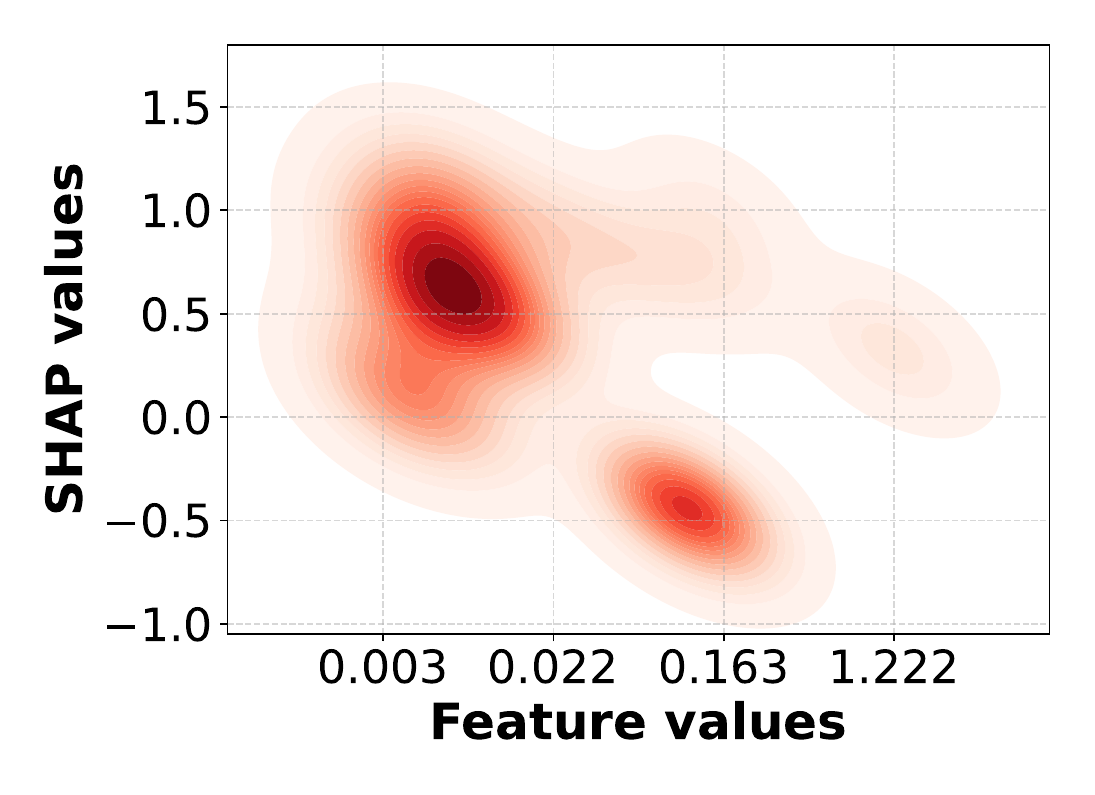}
       \caption{unavail\_second}
        \label{fig:kde_rural_uva}
    \end{subfigure}
    \hfill
    \begin{subfigure}[t]{0.24\linewidth}
        \centering
        \includegraphics[width=\linewidth]{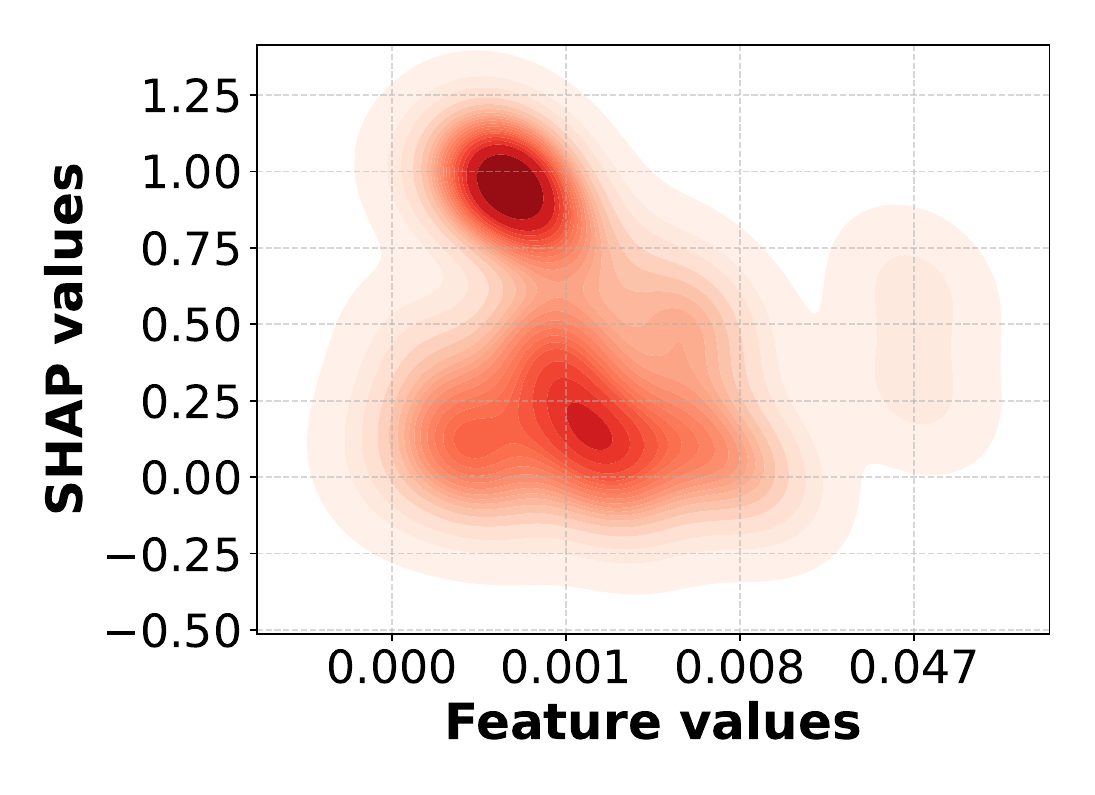}
        \caption{bbe }
        \label{fig:kde_rural_bbe}
    \end{subfigure}
    \hfill
    \begin{subfigure}[t]{0.24\linewidth}
        \centering
        \includegraphics[width=\linewidth]{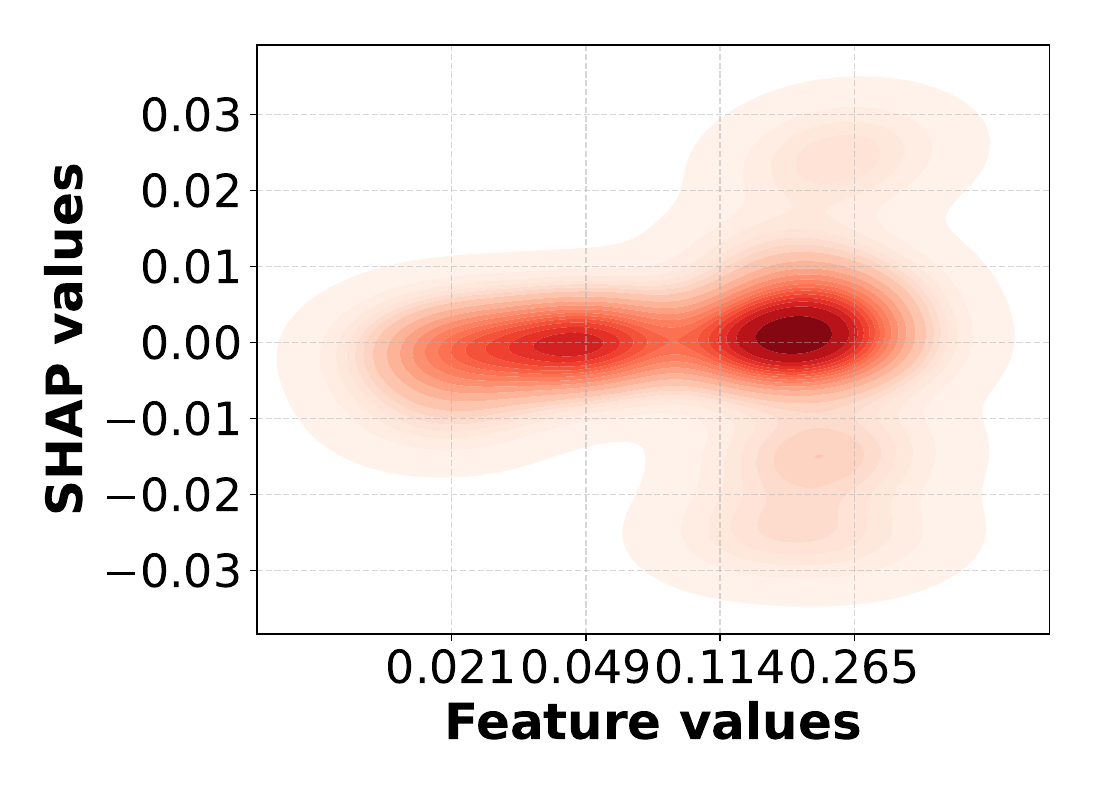}
        \caption{precipitation}
        \label{fig:kde_rural_prec}
    \end{subfigure}
     \begin{subfigure}[t]{0.24\linewidth}
        \centering
        \includegraphics[width=\linewidth]{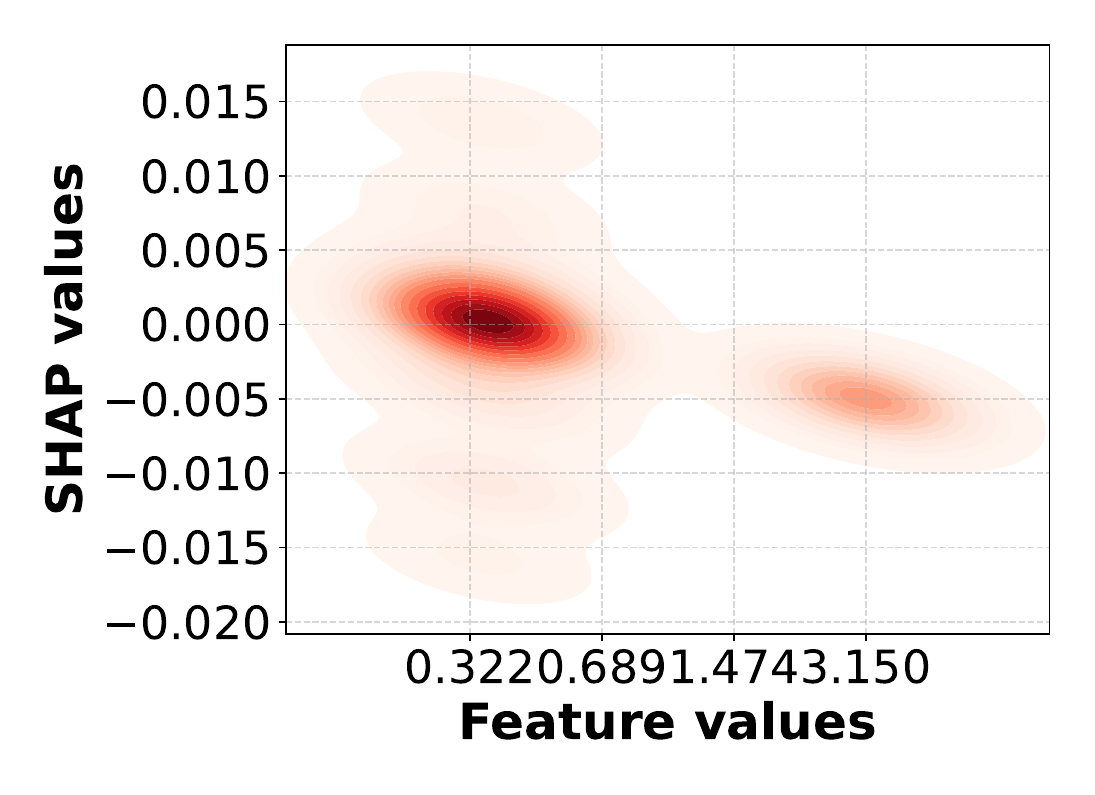}
        \caption{temperature}
        \label{fig:kde_rural_prec}
    \end{subfigure}

    \begin{subfigure}[t]{0.24\linewidth}
        \centering
        \includegraphics[width=\linewidth]{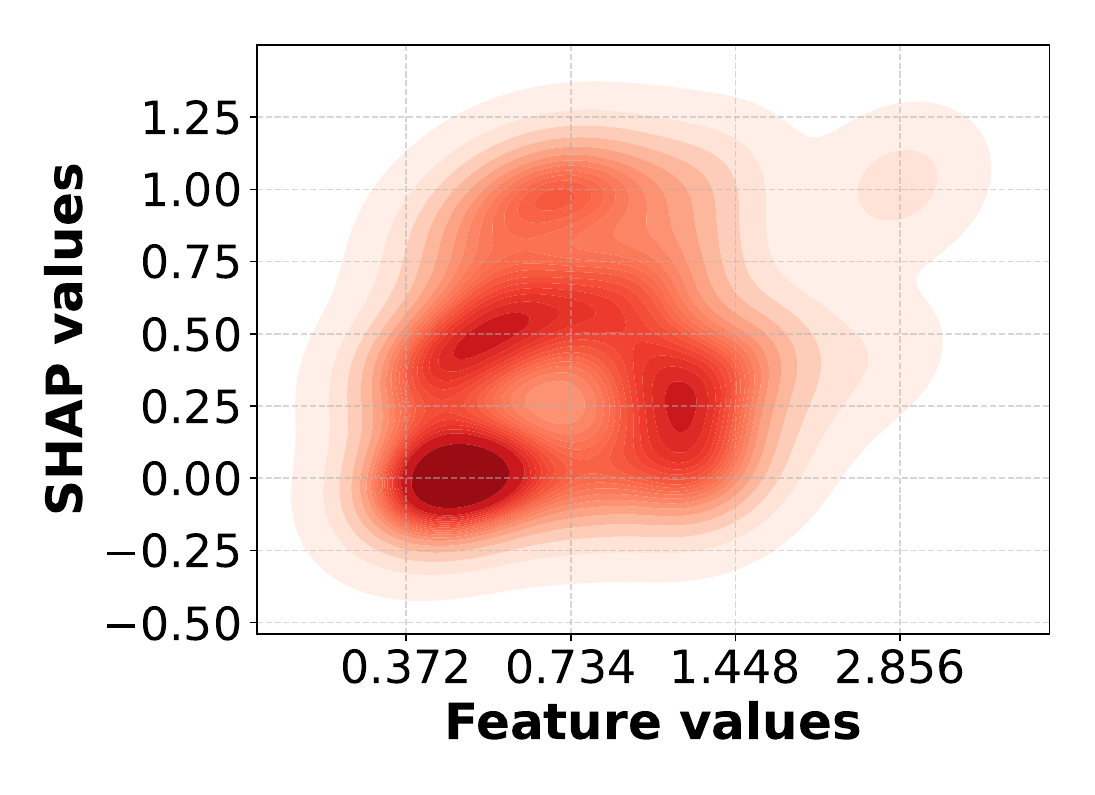}
        \caption{unavail\_second}
        \label{fig:kde_rural_uva}
    \end{subfigure}
    \hfill
    \begin{subfigure}[t]{0.24\linewidth}
        \centering
        \includegraphics[width=\linewidth]{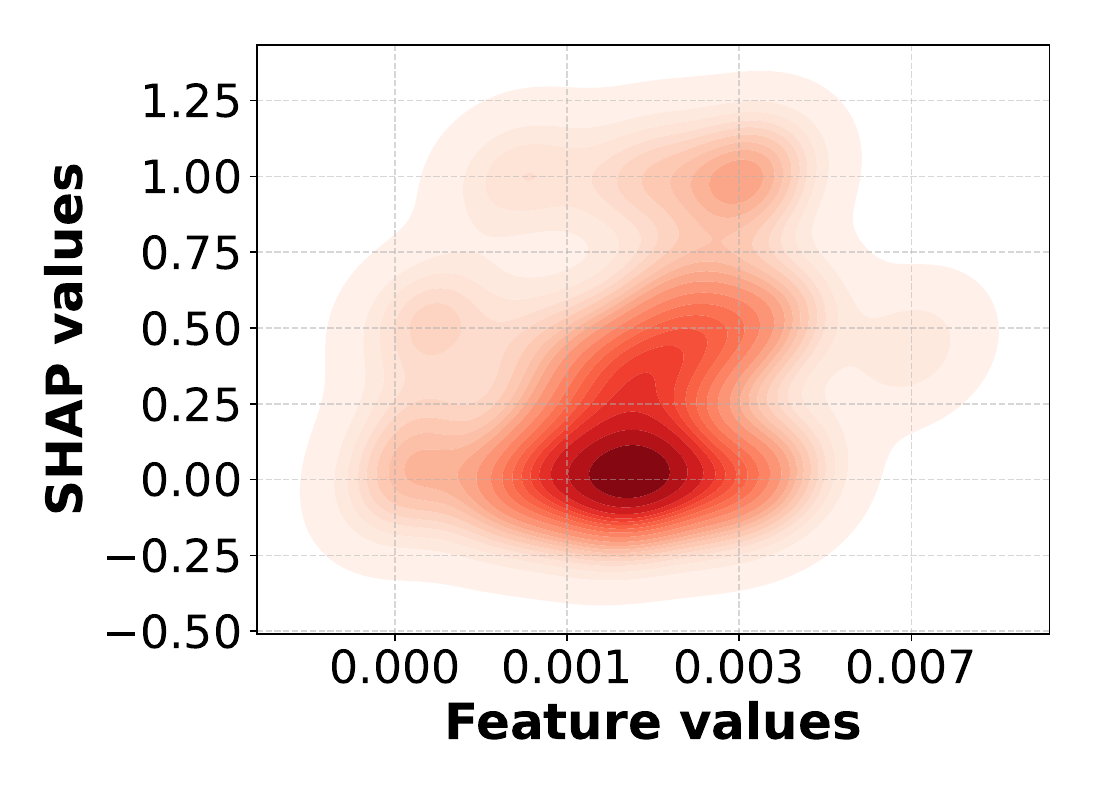}
         \caption{bbe }
        \label{fig:kde_rural_bbe}
    \end{subfigure}
    \hfill
    \begin{subfigure}[t]{0.24\linewidth}
        \centering
        \includegraphics[width=\linewidth]{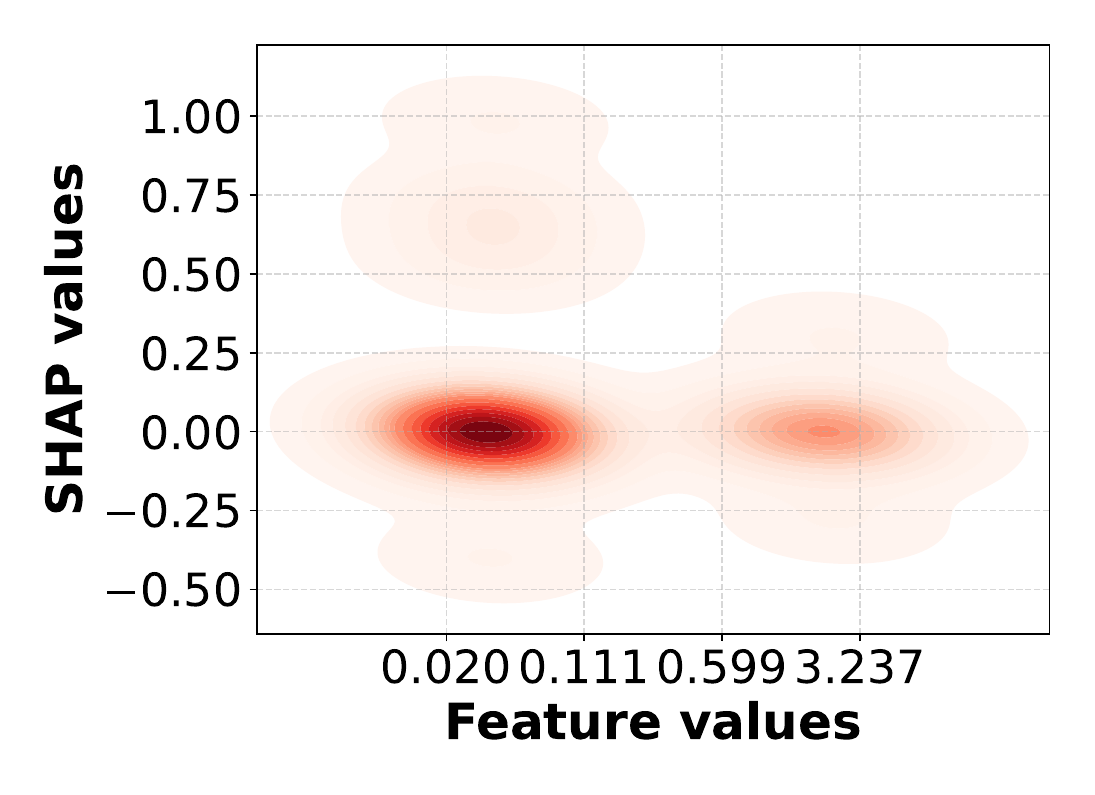}
     \caption{precipitation}
        \label{fig:kde_rural_prec}
    \end{subfigure}
     \begin{subfigure}[t]{0.24\linewidth}
        \centering
        \includegraphics[width=\linewidth]{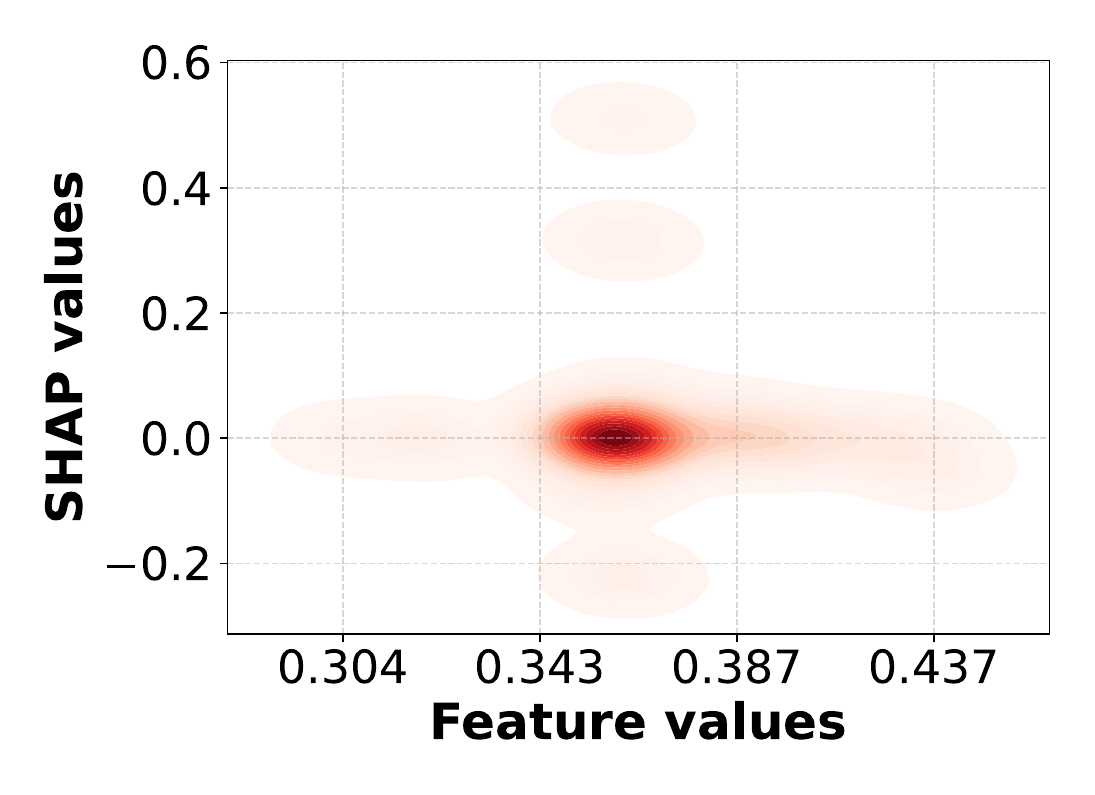}
         \caption{temperature}
        \label{fig:kde_rural_prec}
    \end{subfigure}

\caption{SHAP distributions for the RL and WS features ((a) $\sim$ (d) for rural and (e) $\sim$ (f) for urban) obtained by the \framework using the Gentrap model. Darker colors represent higher density, and lighter colors represent lower density (fewer data samples).}
    \label{fig:kde_uva}
\end{figure*}

Thus, in this second set of evaluation, we present in Figure~\ref{fig:kde_uva} the distribution of important features and SHAP values of the GNN-Transformer model in rural and urban deployments. Specifically, we consider the two most influential features from the RL and weather KPIs, which include \texttt{unavail\_second} and \texttt{bbe} and precipitation and temperature (these two are chosen following \cite{hasan2023transformer, selvaganesh20245g}). For simplicity, we present the distribution of features in the last time step where we observe maximum changes. We see that for both urban and rural scenarios, \texttt{unavail\_second} and \texttt{bbe} have high SHAP values in most samples, indicating a stronger influence on model output. 
Furthermore, these high SHAP values are clustered together to indicate that the model output will vary proportional to these two key KPIs. However, the values of two weather features spread across samples with no specific patterns in the RLFs, i.e., they have a minimal impact on the predictions. 
Finally, we identified another interesting distinction between urban and rural deployment. \texttt{unavail\_second} has different levels of contribution to urban and rural deployments. After a certain threshold, its contribution in the rural scenario is not significant, which is not the case in the urban scenario. We suspect that this behavior may have an influence on the false positive case in urban deployment leading to a slight performance degradation.

\begin{figure*} [ht]
    \centering   
\begin{subfigure}{0.495\textwidth}
        \centering
        \includegraphics[width=.88\linewidth]{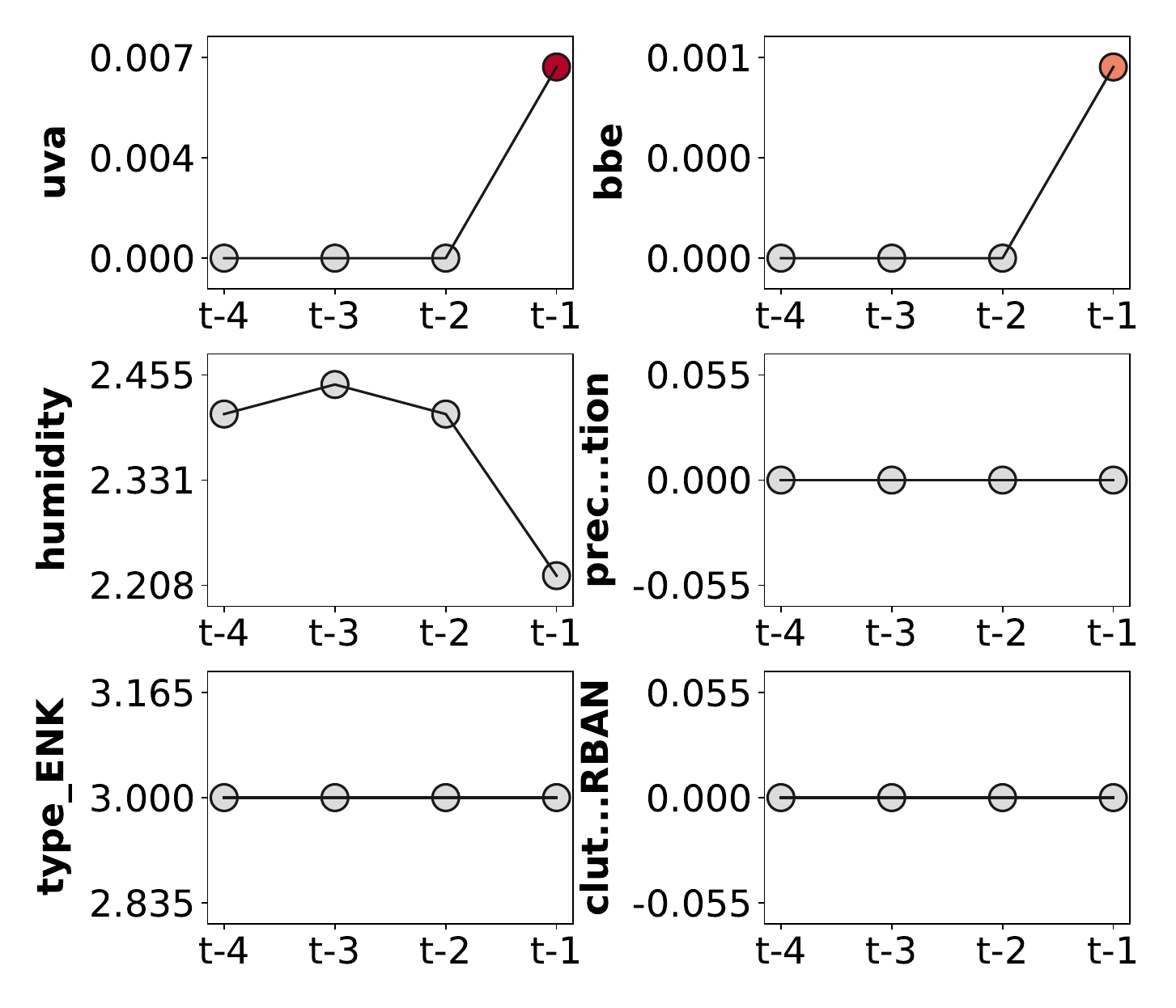}
        \caption{Example 1}
        \label{fig:sub1}
    \end{subfigure}
    \hfill
    \begin{subfigure}{0.495\textwidth}
        \centering
        \includegraphics[width=.88\linewidth]{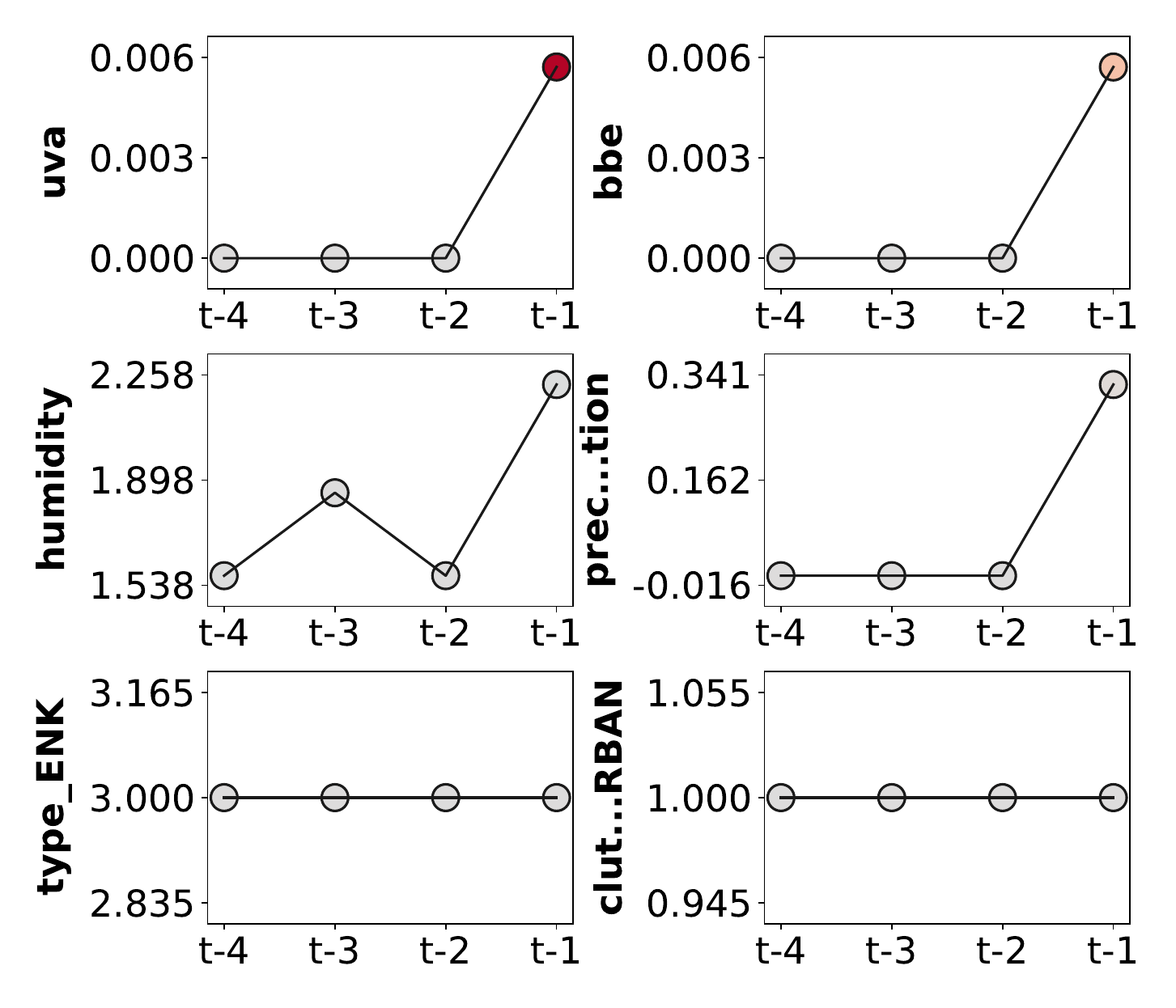}
        \caption{Example 2}
        \label{fig:sub2}
    \end{subfigure}
    
    \caption{The SHAP values for some features for different time steps of failure prediction obtained by the \framework predicted in the GNN-Transformer model .}
    \label{fig:input_exp}
\end{figure*}

Figure~\ref{fig:input_exp} presents the next evaluation results on changes in features over time and how \framework captures that using two RL, WS, and static features. The color of the circles at each step provides the SHAP intensity (red for positive SHAP values; blue for negative SHAP values). We see that at the $t-1$ time step (the immediate previous days), the values of \texttt{unavail\_second} and \texttt{bbe} increase to a higher value, which the model gives more importance for the failure prediction, and \framework can retrieve it successfully (indicated by red circles). Additionally, the WS features also change over time, but our framework does not place much importance on them, as those values are changing randomly rather than having a sequential pattern (e.g., in sample 1, humidity is decreasing, but precipitation is unchanged; in sample 2, humidity and precipitation are increasing) that could contribute to a failure. Similarly, static features do not contribute to failure prediction as they remain similar over time, and \framework also does not give much importance to them.

We have conducted a similar analysis for the LSTM+ model and present the outcome in Figure \ref{fig:failure_sum_lstm}. The results show a similar trend to that of GenTrap, i.e., \texttt{unavail\_second} and \texttt{bbe} are of highest importance. However, LSTM+ also gives importance to the WS features (e.g., temperature, precipitation, and humidity) for failure predictions, suggesting that the model architecture has influence on the contextual feature importance. Specifically, though LSTM-based models are suitable for sequential data, they struggle with complex time series involving multivariate features. Consequently, LSTM-based models may ineffectively learn feature importance compared to the Transformer-based model and suffer from a lower F1-score.

In summary, we conclude that static features do not contribute to either model in the failure predictions and can be removed for simplification. WS features are not contributing to the GNN-Transformer model prediction; however, the LSTM-based model uses some WS time-series features, but due to its architectural limitations, the model offers lower performances.

\begin{table*} 
\centering
\footnotesize
\renewcommand{\arraystretch}{1.5}
\caption{Performance comparison of all RLF models. Higher values correspond to better performance. The best results are in \textbf{bold}.}
\label{tb:result}  
\begin{tabular}{l|ccc|ccc} 
  \toprule
  \multicolumn{1}{c}{} & \multicolumn{3}{c}{\textbf{Rural}} &  \multicolumn{3}{c}{\textbf{Urban}} \\
  \midrule
  Model & Precision & Recall & F1-Score & Precision & Recall & F1-Score \\ 
  \midrule
  \rowcolor{gray!30}
\textbf{LTrans} & \textbf{.89 $\pm$ .03} & \textbf{.98 $\pm$ .01} & \textbf{.94 $\pm$ .02}  & \textbf{.73 $\pm$ .05} & \textbf{.97 $\pm$ .03} & \textbf{.83 $\pm$ .03} \\
\textbf{Gentrap} & .88 $\pm$ .06 & .92 $\pm$ .07 & .90 $\pm$ .03 & .58 $\pm$ .06 & .88 $\pm$ .10 & .69 $\pm$ .05 \\ 
\rowcolor{gray!30}
\textbf{LLSTM+} & .46 $\pm$ .19 & .92 $\pm$ .05 & .59 $\pm$ .18 & .18 $\pm$ .05 & .88 $\pm$ .20 & .29 $\pm$ .07 \\
\textbf{LSTM+} & .41 $\pm$ .18 & .70 $\pm$ .11 & .52 $\pm$ .16 & .16 $\pm$ .08 & .65 $\pm$ .17 & .25 $\pm$ .12 \\ 
\bottomrule
\end{tabular}
\end{table*}

\subsection{Are lightweight models comparable to the black-box models? (RQ2)}

This section compares the performance of existing models with their simplified versions to address RQ2. Based on the findings of RQ1, we derive new lightweight models. We evaluate the performance of these models against the original models in both urban and rural datasets. Table \ref{tb:result} summarizes these results (averaged across all data folds F0 $\sim$ F4). 
In general, the \newmodel Transformer-based model (LTrans) offers the best F1-score, which is 4\% and 20\% better than GenTrap in rural and urban settings, respectively.  Thus, LTrans may be more effective in dense urban deployment than in rural deployment. Moreover, the consistently high (around 98\%) recall values demonstrate that LTrans can effectively predict almost all failures. This superior performance of LTrans comes from the fact that the Transformer can focus on the most relevant features by reducing data redundancy through feature selection and retaining only useful RL KPI features.

We observe a similar trend for the LSTM+ and its lightweight version (LLSTM+), although these models generally perform worse than the Transformer-based ones. In particular, LLSTM+ offers 13\% better F1-score than the original LSTM+ model in the rural deployment.  This gain results from the fact that LLSTM+ emphasizes more effectively on important features during pattern learning. In contrast, the improvement in the urban scenario is negligible. This occurs because the LLSTM+ model encompasses both RL and WS features similar to the original model, which may not be effective in the dense urban scenario. LSTM is less suitable for handling multivariate and complex time series data at the scale. Consequently, the feature selection for the LSTM-based model is less accurate, resulting in a weaker performance in its lightweight version compared to the Transformer-based model.

\begin{figure} [h]
    \centering   \includegraphics[width=0.5\textwidth, clip, trim=0 35 0 0]{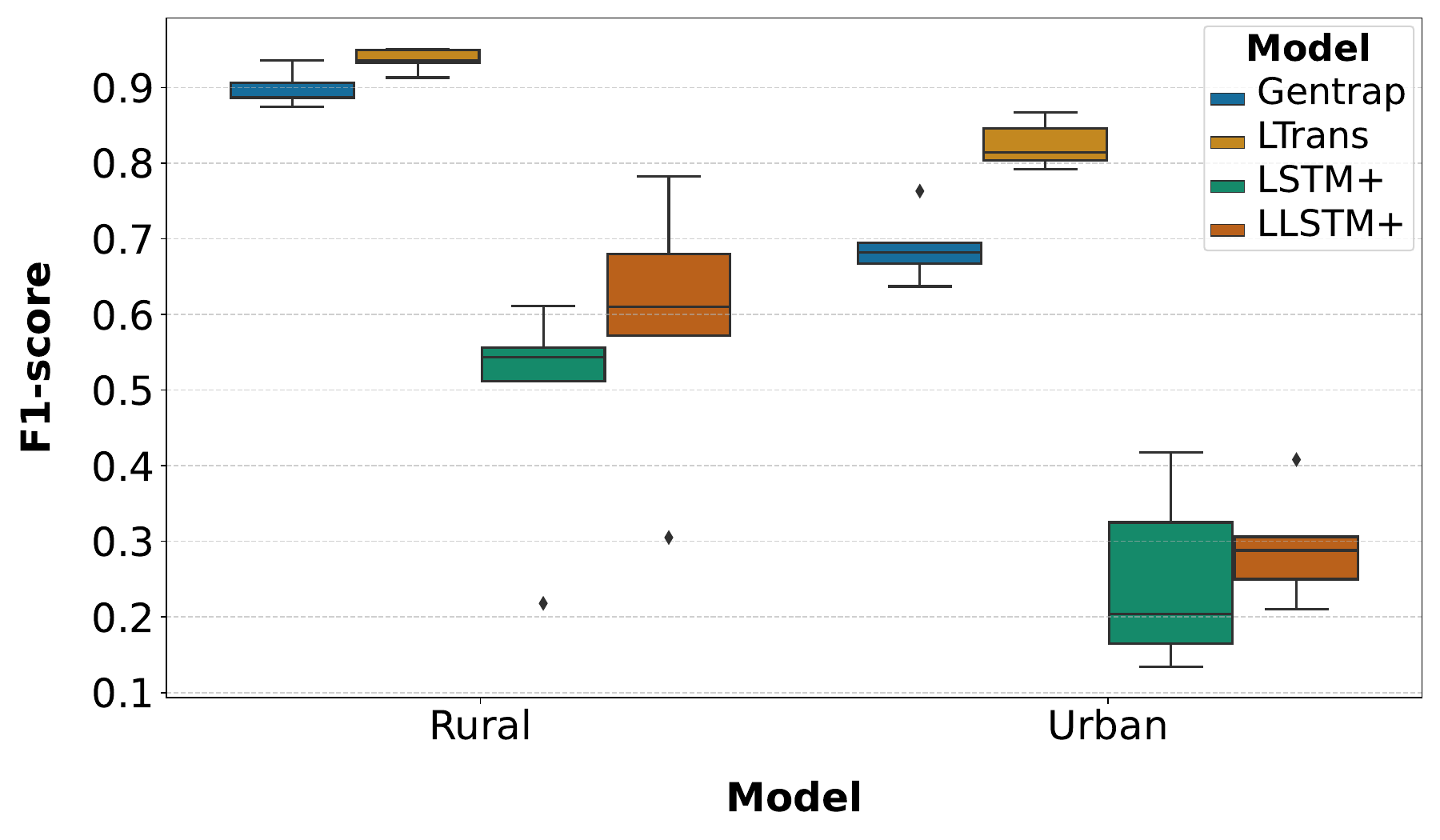}
    \caption{Box-plots of the distribution of F1-score of the evaluated models across the five data folds.}
    \label{fig:boxplotf1}
\end{figure}

To better assess the lightweight models compared to the existing ones, we present F1-score box-plots in Figure~\ref{fig:boxplotf1}. We see that the variability of the Transformer-based models is similar for rural deployments. However, LTrans has the highest median F1-score and a smaller spread than GenTrap. Although LTrans exhibits a wider spread for urban deployment, it still outperforms all other models. For the LSTM-based models, the median F1-score improves over the existing one; however, the variability is noticeably higher. Both LSTM models show larger spreads and potential outliers compared to the Transformer-based models. This also demonstrates the superiority of the Transformer-based models over their LSTM counterparts.

\subsection{Which features contribute to the RLF prediction, and why? (RQ3)}

\begin{figure}[h]
\centering
\begin{subfigure}[t]{0.45\textwidth}
    \centering
    \includegraphics[width=.8\textwidth]{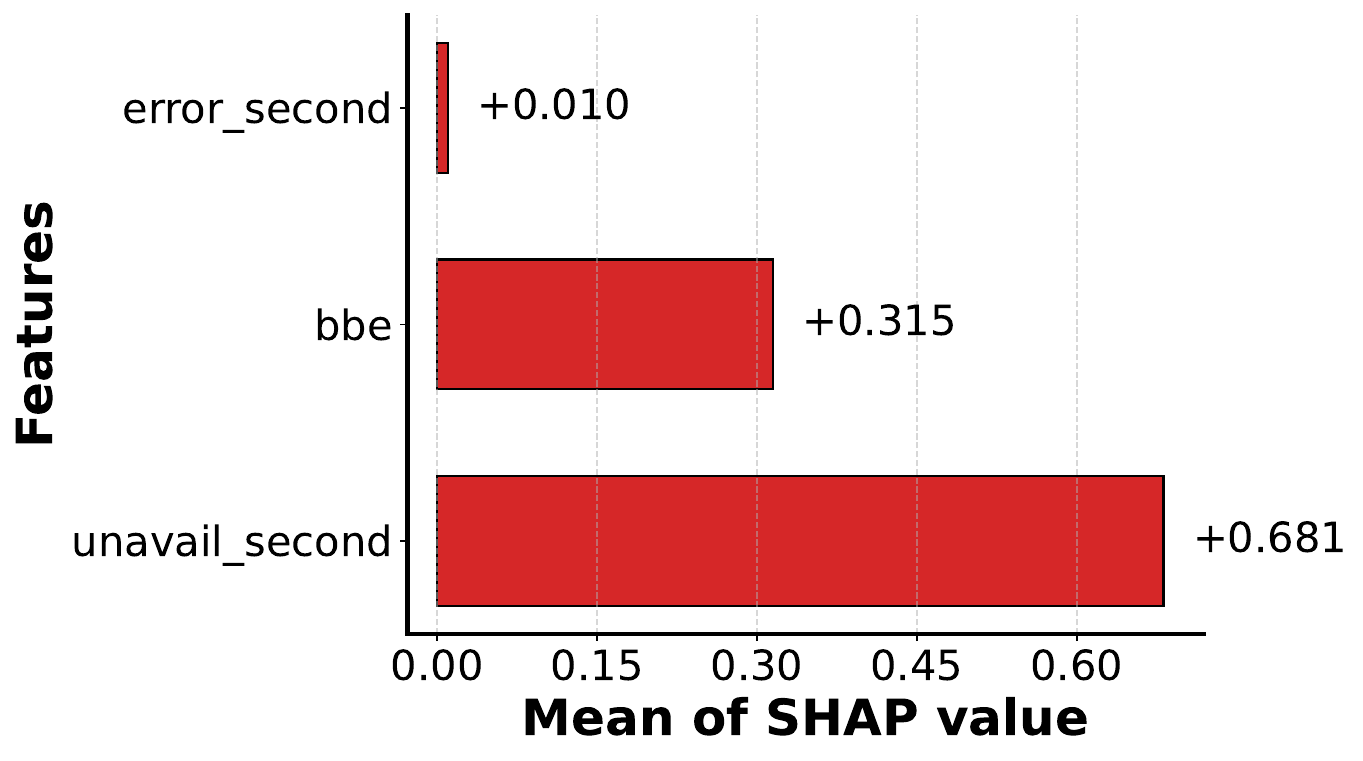}
    \caption{Rural}
    \label{fig:exp_rural}
\end{subfigure}
\hfill
\begin{subfigure}[t]{0.45\textwidth}
    \centering
    \includegraphics[width=.8\textwidth]{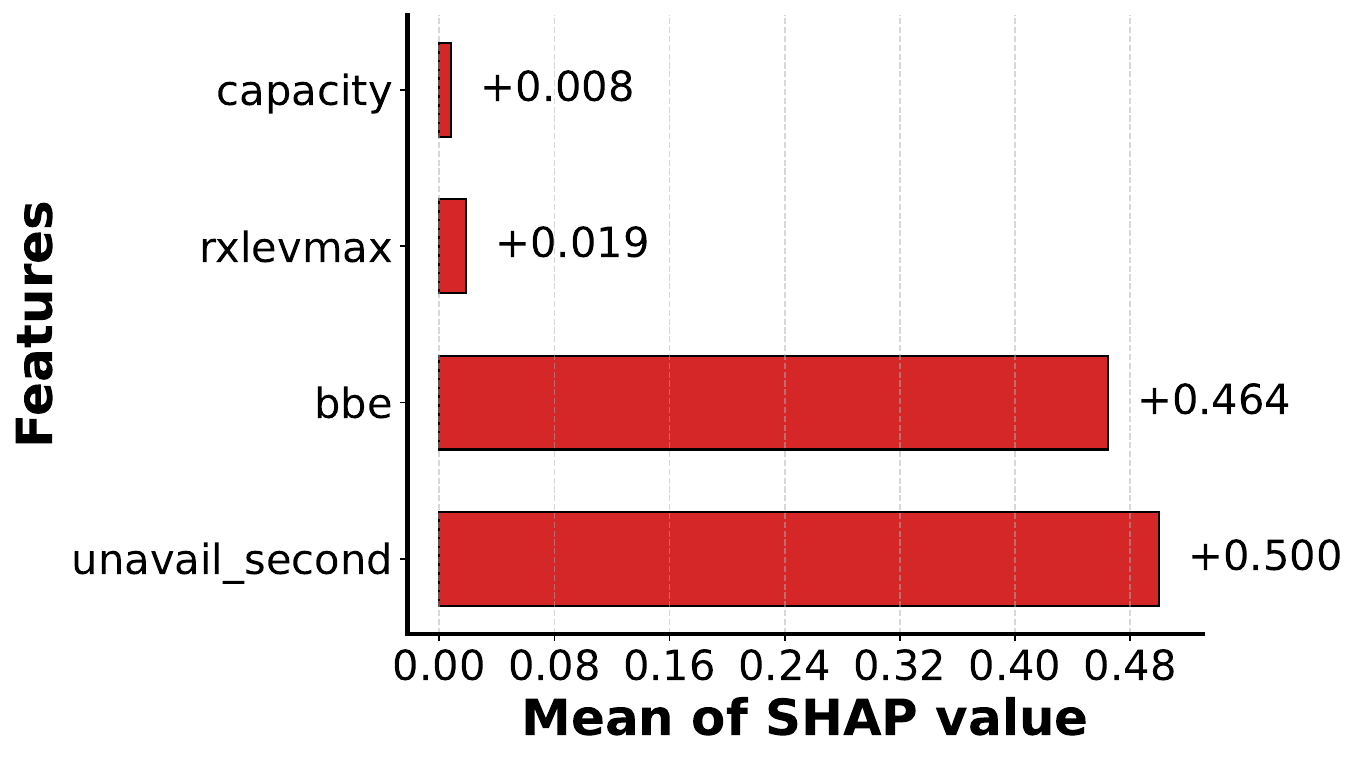}
    \caption{Urban}
    \label{fig:exp_urban}
\end{subfigure}
\caption{The mean SHAP values across all failures.}
\label{fig:exp_shap_comparison}
\end{figure}

We identified that \texttt{unavail\_second} and \texttt{bbe} are the most influential features for RLF prediction, along with few other RL features, while WS and static features contribute minimally. In particular, the Transformer model with only RL features offers the best performance. Thus, in this final evaluation, we consider LTrans to answer RQ3.  In this case, we generate the feature importance of LTrans using \framework, which are illustrated in Figure~\ref{fig:exp_rural} and Figure\ref{fig:exp_urban}. We observe that \texttt{unavail\_second} and \texttt{bbe} are indeed the most important features. Additionally, \texttt{error\_second}, \texttt{rxlevmax}, and \texttt{capacity} have a contribution to the failure predictions. 
According to the literature, \texttt{unavail\_second} and \texttt{bbe} are strongly associated with RLF and can cause failures when they exceed specific thresholds \cite{3gpp-ts-36.331, itu-r-f557, hasan2023transformer}, which our framework successfully captures in both deployments. 
Furthermore, link capacity reduction and increased error seconds can also lead to link failures \cite{libunao2017autonomic, ITUTG826, 3gpp-ts-36.331}; however, these occurrences are infrequent in the datasets and are assigned lower importance.

\begin{figure*}[ht]
\centering
\begin{subfigure}[t]{0.475\textwidth}
    \centering
    \includegraphics[width=0.75\textwidth]{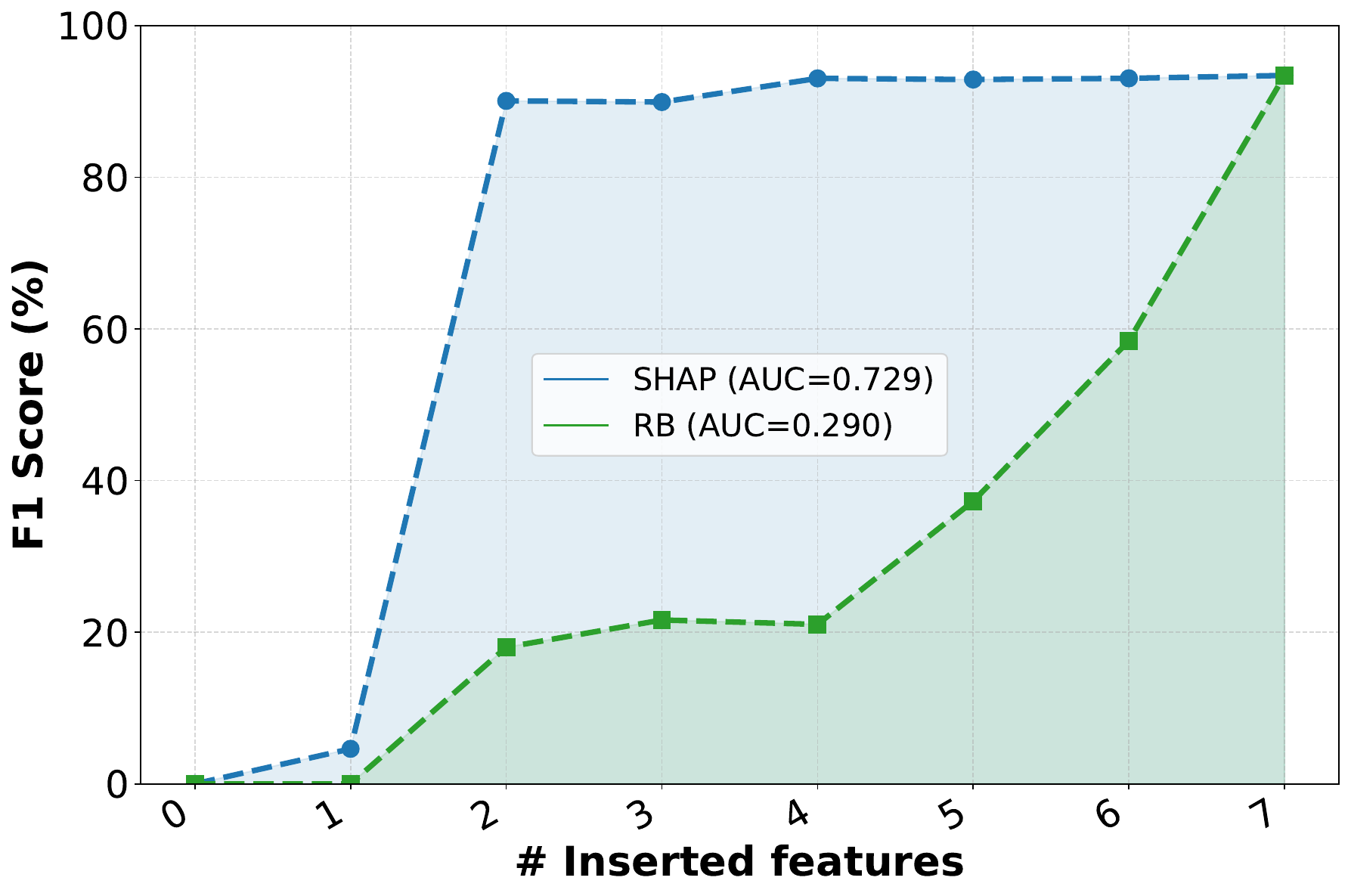}
    \caption{Rural}
    \label{fig:insertionr}
\end{subfigure}
\hfill
\begin{subfigure}[t]{0.475\textwidth}
    \centering
    \includegraphics[width=0.75\textwidth]{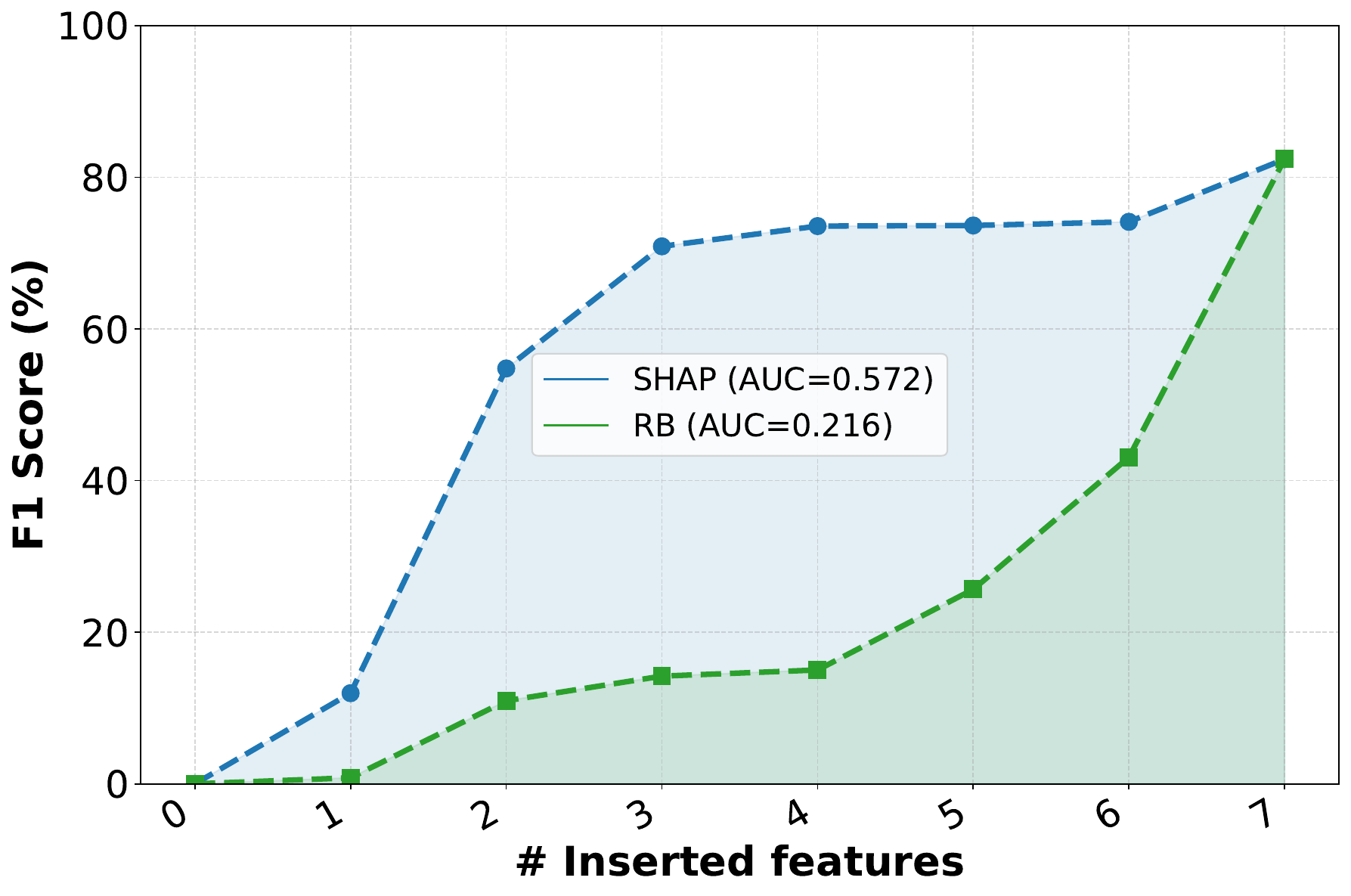}
    \caption{Urban}
    \label{fig:insertionu}
\end{subfigure}
\caption{Insertion test for validating feature importance, using SHAP (blue) and a RANDOM baseline (green). Higher AUC score indicates that SHAP correctly identified the most important features.}
\label{fig:insertion_comparison}
\end{figure*}

\begin{figure*}[ht]
\centering
\begin{subfigure}[t]{0.475\textwidth}
    \centering
    \includegraphics[width=0.75\textwidth]{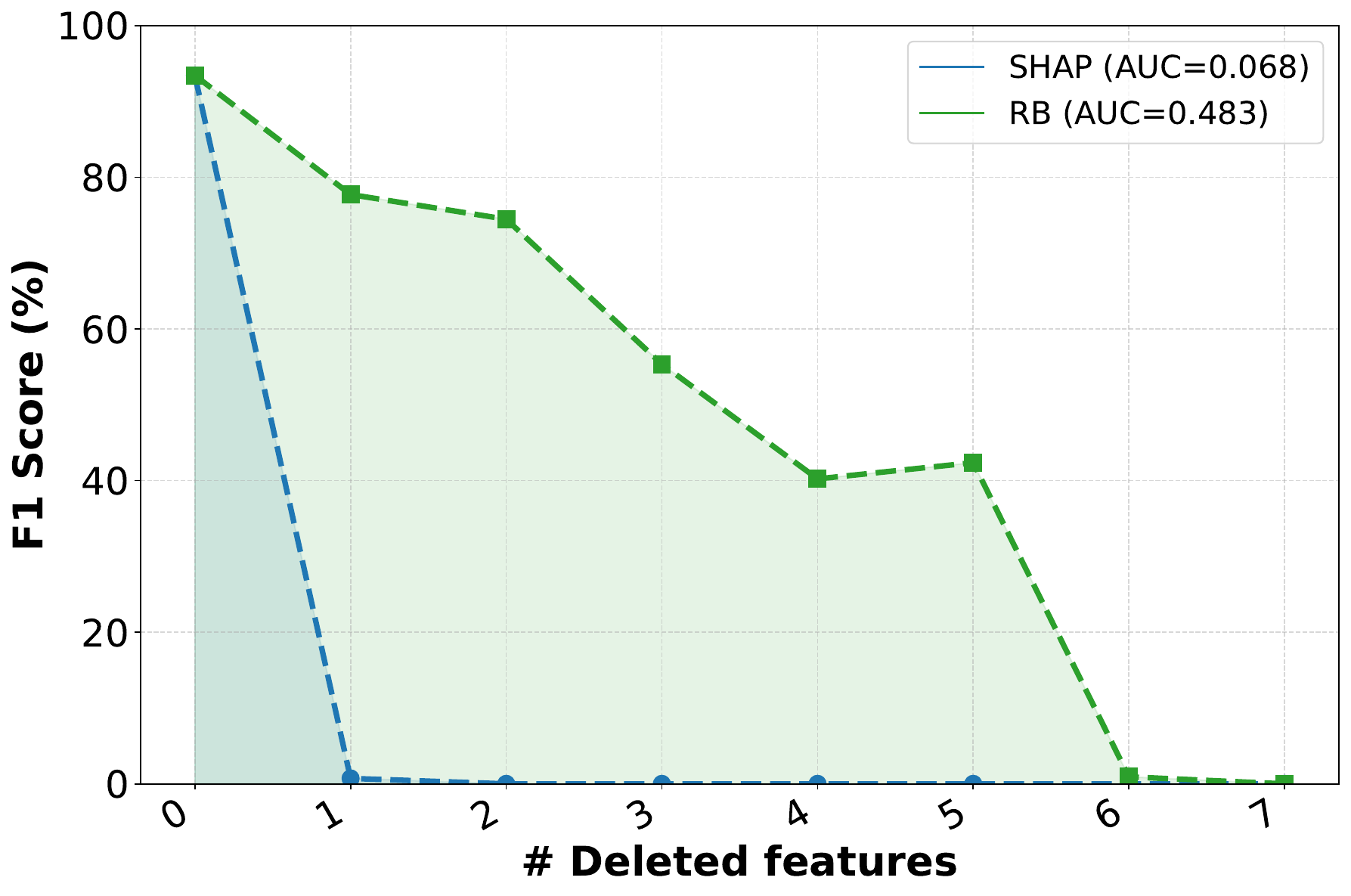}
    \caption{Rural}
    \label{fig:deletion}
\end{subfigure}
\hfill
\begin{subfigure}[t]{0.475\textwidth}
    \centering
    \includegraphics[width=0.75\textwidth]{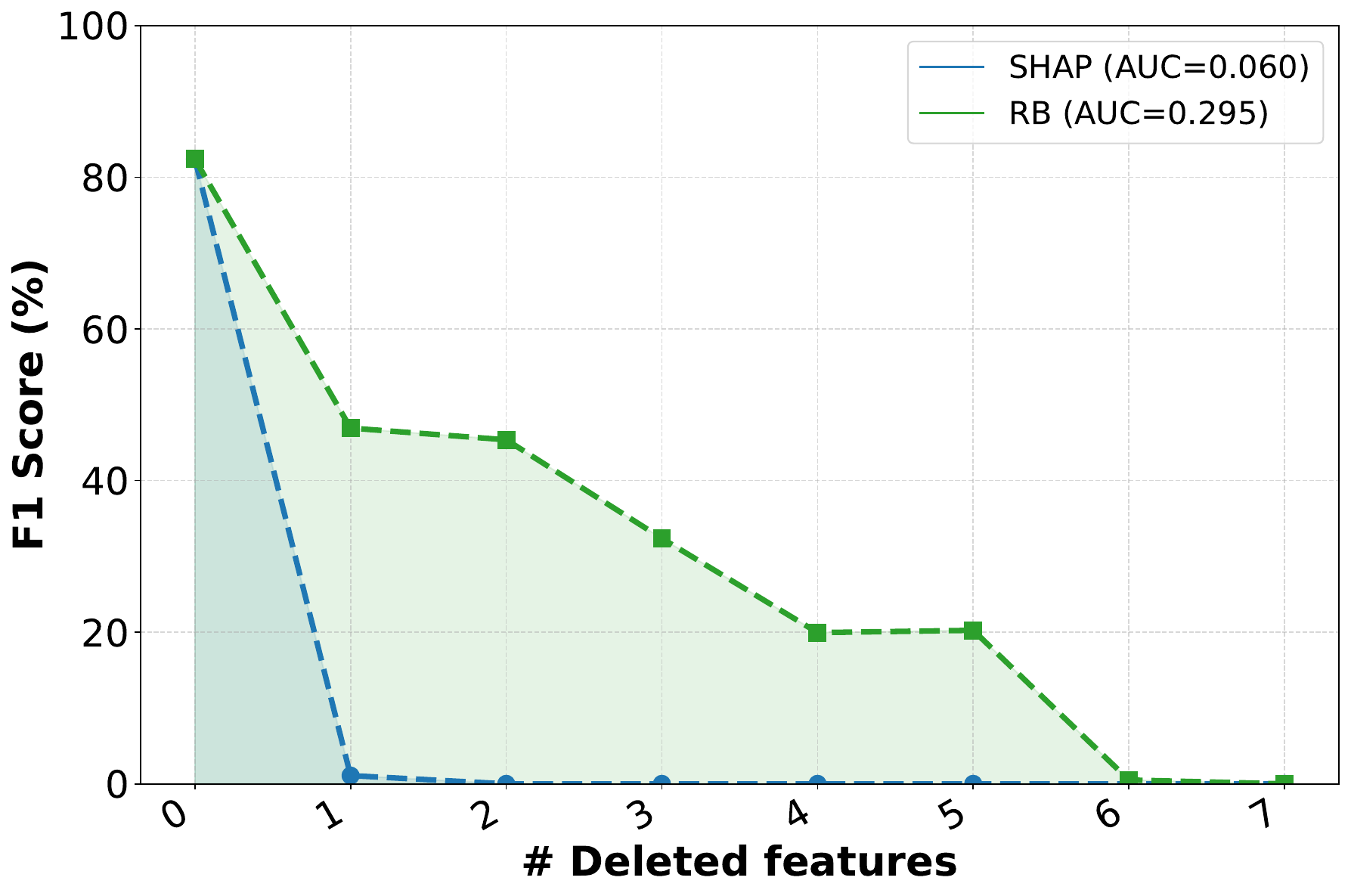}
    \caption{Urban}
    \label{fig:deletion}
\end{subfigure}
\caption{Deletion test for validating feature importance using SHAP (blue) and a RANDOM baseline (green). A lower AUC score indicates that SHAP correctly identified the most important features.}
\label{fig:deletion_comparison}
\end{figure*}

In the second set of evaluations, we conduct widely adopted insertion and deletion tests of feature attribution to identify the most influential features for a given model \cite{hama2023deletion}. Figure~\ref{fig:insertionr} and Figure~\ref{fig:insertionu} present the insertion test results of the LTrans model in the rural and urban settings. In this test, features are ranked by their importance and gradually inserted, starting from an empty set. The results show that inserting only the two top features yields 96\% of the total F1-score in the rural and 67\% in the urban deployment. This indicates that the top features explain most of the model performance in rural scenarios, while additional features contribute modestly in urban settings. The area under the curve (AUC) exceeds $0.73$ in rural and $0.57$ in urban, both outperforming a random baseline (RB), confirming the reliability of the explanation ability of \framework. We observe a similar performance trend in the case of the deletion test (Figure \ref{fig:deletion_comparison}), i.e., removing the most important feature (\texttt{unavail\_second}) causes the F1-score to drop to zero. Removing the second most important feature (and \texttt{bbe}) results in the F1-score reaching zero for both datasets, demonstrating that these two features are the key drivers of failure predictions.

In summary, \framework effectively identifies the key causes of radio link failures, aligning with established findings in telecommunication studies and supported by evidence that ensures the validity and reliability of the results. By quantifying the contribution of critical features and linking them to known failure mechanisms, the framework improves both the interpretability and trustworthiness of neural network models. Consequently, it enables a clearer understanding of the underlying problem while delivering a more transparent and higher-performing predictive model.

\section{Conclusion}
\label{sec:conclusions}

This work proposed \framework, an explainable framework for failure prediction in 5G RAN that enhances reliability and transparency for operators, enabling more effective proactive mitigation measures. The framework not only identifies the key features responsible for failures but also provides interpretable reasoning to support decision-making. By pruning less relevant features, it simplifies data management and enables lightweight yet high-performing models. Our results demonstrate that incorporating explainability improves both interpretability and predictive accuracy.

While the dataset used well-established and commonly used in the research community on 5G, it contains time series that are relatively short. Testing the framework on longer time series would allow us to better assess its robustness and generalization to more complex temporal dynamics. Additionally, the current explanation method focuses on feature importance, which has proven effective in identifying critical failure causes. However, integrating counterfactual explanations could further enhance the interpretability by showing how minimal changes in features might alter model predictions. As part of future work, we plan to address these limitations by benchmarking \framework on longer time series and more complex datasets. We also aim to integrate counterfactual explanation techniques to complement feature-based explanations, providing deeper insights for operational decision-making. Finally, we intend to develop instinctive model simplification capabilities to enhance the adaptability of the framework across a broader range of network management problems.



\bibliographystyle{IEEEtran}
\bibliography{sn-bibliography}

\end{document}